\documentclass[journal]{IEEEtran}
\usepackage{pifont}
\usepackage{times,amsmath,color,amssymb,graphicx,epsfig,cite,psfrag,subfigure}
\usepackage{mathrsfs}
\usepackage{amsfonts}
\long\def\symbolfootnote[#1]#2{\begingroup%
\def\thefootnote{\fnsymbol{footnote}}\footnote[#1]{#2}\endgroup}

\begin{document}
\title{Mobile Data Offloading through A Third-Party WiFi Access Point: An Operator's Perspective\vspace{-1mm}}
%\author{\authorblockN{Xin Kang, Yeow-Khiang Chia, and Sumei
%Sun}
%\authorblockA{
%Institute for Infocomm Research, 1 Fusionopolis Way,
%$\sharp$21-01 Connexis, South Tower, Singapore 138632\\
%Email: \{xkang, chiayk, sunsm\}@i2r.a-star.edu.sg}\vspace{-6mm}}
%\maketitle
\author{~~~~~~~~~~~~~~Xin~Kang,~\IEEEmembership{Member,~IEEE}, ~Yeow-Khiang
Chia,~\IEEEmembership{Member,~IEEE}, ~\newline Sumei
Sun,~\IEEEmembership{Senior Member,~IEEE},~and Hon Fah
Chong,~\IEEEmembership{Member,~IEEE}
\thanks{X. Kang, Y.-K Chia, S. Sun, and H. F. Chong are with Institute for Infocomm Research, 1 Fusionopolis Way,
$\#$21-01 Connexis, South Tower, Singapore 138632 (E-mail: \{xkang,
chiayk, sunsm, hfchong\}@i2r.a-star.edu.sg).} } \maketitle

\begin{abstract}
\textcolor[rgb]{0,0,0}{WiFi offloading is regarded as one of the
most promising techniques to deal with the explosive data increase
in cellular networks due to its high data transmission rate and low
requirement on devices. In this paper, we investigate the mobile
data offloading problem through a third-party WiFi access point (AP)
for a cellular mobile system. From the cellular operator's
perspective, by assuming a usage-based charging model, we formulate
the problem as a utility maximization problem. In particular, we
consider three scenarios: (i) successive interference cancellation
(SIC) available at both the base station (BS) and the AP; (ii) SIC
available at neither the BS nor the AP; (iii) SIC available at only
the BS. For (i), we show that the utility maximization problem can
be solved by considering its relaxation problem, and we prove that
our proposed data offloading scheme is near-optimal when the number
of users is large. For (ii), we prove that with high probability the
optimal solution is One-One-Association, i.e., one user connects to
the BS and one user connects to the AP. For (iii), we show that with
high probability there is at most one user connecting to the AP, and
all the other users connect to the BS. By comparing these three
scenarios, we prove that SIC decoders help the cellular operator
maximize its utility. To relieve the computational burden of the BS,
we propose a threshold-based distributed data offloading scheme. We
show that the proposed distributed scheme performs well if the
threshold is properly chosen.}
\end{abstract}

\begin{keywords}
WiFi offloading, utility maximization, user association, integer
programming, schur convex.
\end{keywords}

%\newpage
\section{Introduction}\label{Sec-intro}
\textcolor[rgb]{0,0,0}{The rapid development of mobile phones and
mobile internet services in recent years has generated a lot of data
usage over the cellular network\cite{Cisco}. The unprecedented
explosion of mobile data traffic has led to overloaded cellular
networks. For example, in metro areas and during peak hours, most 3G
networks are overloaded\cite{Wavion}. Mobile users in overloaded
areas will have to experience degraded cellular services, such as
low data transmission rate and low quality phone calls.}

\textcolor[rgb]{0,0,0}{A straightforward approach to address the
above problem is to upgrade the  cellular network to the more
advanced 4G network. Another approach is to deploy more base
stations (BSs) with smaller cell size such as
femtocells\cite{VChan-Sep2008,XKangJSAC2011}. However, these
approaches incur increase in infrastructure cost. A more
cost-effective approach is to offload some of the mobile traffic to
WiFi networks, which is often referred to as WiFi offloading. It has
a few advantages: (i). No user equipment upgrading is required. This
is because most of the mobile data services are created by
smartphones which already have  built-in WiFi modules. (ii). No
licensed spectrum is required. WiFi devices operate in unlicensed
and world-unified 2.4GHz and 5GHz bands. (iii). High data rates.
IEEE 802.11n WiFi can deliver data rates as high as 600Mbps and IEEE
802.11ac can deliver up to 6.933Gbps \cite{CiscoWiFi}, which is much
faster than 3G. (iv). Low infrastructure cost. The WiFi routers are
much cheaper than the cellular BSs.}

For the aforementioned reasons, WiFi offloading becomes a hot
research topic and has attracted the attention of many researchers
all over the world \cite{ABalasubramanian}--\kern-1mm\cite{JLeeSDP}.
The feasibility of augmenting 3G using WiFi was investigated in
\cite{ABalasubramanian}. The performance of 3G mobile data
offloading through WiFi networks for metropolitan areas was studied
in \cite{KLee}. The numbers of APs needed for WiFi offloading in
large metropolitan area was studied in \cite{SDimatteocy}. Different
approaches to implement WiFi offloading and to improve the
performance of WiFi offloading were investigated in
\cite{Sankaran}--\kern-1mm\cite{niyato}. The load-balancing and
user-association problem for offloading in heterogeneous networks
with cellular networks and small cells are investigated in
\cite{Jeff0}--\kern-0.6mm\cite{CKHo}.  In \cite{Jeff0}, the authors
investigated the outage probability and ergodic rate when a flexible
cell association scheme is adopted. In \cite{Jeff2}, the authors
developed a general and tractable model for data offloading in
heterogeneous networks with different tiers of APs.  In
\cite{Jeff1}, the authors investigated the downlink user association
problem for load balancing in a heterogeneous cellular networks. In
\cite{CKHo}, the authors investigated the data offloading schemes
for load coupled networks, and showed that the optimal loading is
tractable when proportional fairness is considered. Recent works
\cite{LGao}--\kern-1mm\cite{5} investigated the network economics of
data offloading through WiFi APs using game theory
\cite{GameTheory1993}.

\textcolor[rgb]{0,0,0}{Different from the above work, in this paper,
we consider the scenario that there is a third-party WiFi AP
providing data offloading service with a usage-based charging
policy. We investigate the data offloading problem through such a
third-party WiFi AP for a cellular mobile communication system. From
business perspective, the cellular operator aims to maximize its
revenue. Thus, in this paper, we investigate the data offloading
problem from the economic point of view. We formulate the problem as
a utility maximization problem and derive the corresponding data
offloading schemes for the cellular operator. In particular, we
consider three scenarios, namely, SIC available at both the BS and
the AP, SIC available at only the BS, and SIC available at neither
the BS nor the AP. We study the  different utility functions and
propose different data offloading schemes.}

%Different from the above work, in this paper, we investigate the
%network economics of the data offloading problem through a
%third-party WiFi AP for a cellular mobile system. From the cellular
%operator's perspective, by assuming the data offloaded through the
%third-party WiFi AP is charged based on usage, we formulate the
%mobile data offloading problem as a utility maximization problem. By
%considering whether the successive interference cancellation (SIC)
%decoders are available at the BS and/or the WiFi AP, we study
%different utility functions and derive the corresponding optimal
%data offloading schemes when the number of users is large.

The main contribution and results of this paper are summarized as
follows.

\begin{itemize}
  \item \emph{SIC available at both the BS and the AP:} The utility maximization problem for this case is solved by considering its
relaxation problem. We show that the relaxation problem is a convex
optimization problem. By using the convex optimization techniques,
we prove that there is at most one user with fractional indicator
function. A data offloading scheme is then obtained by rounding the
fractional indicator function to its nearest integer. It is strictly
proved that the proposed data offloading scheme is near-optimal when
the number of users is large.
  \item \emph{SIC available at neither
the BS nor the AP:} For this case, we rigorously prove that when the
number of users is large, the optimal solution is
One-One-Association, i.e., the user with the best user-to-BS channel
connects
  to the BS and that with the best user-to-AP channel connects to the WiFi AP.
  \item \emph{SIC available at only the BS:} For this
  case, we show that when the number of users is large, there is at most one user
  connecting to the WiFi AP, and all the other users connect to the
  BS. A polynomial-time algorithm is developed to find the optimal offloading
  scheme.
  \item \emph{SIC is beneficial for the cellular
  operator:} We rigorously prove that SIC decoders are
  beneficial for the cellular operator in terms of maximizing its
  utility.
  \item \emph{Distributed data offloading scheme:} We propose a threshold-based distributed data offloading
scheme for the case when SIC decoders are available at both the BS
and the AP. We prove that the proposed distributed scheme can
achieve the same performance as the centralized data offloading
scheme once the threshold is properly chosen.
\end{itemize}

The rest of this paper is organized as follows: In Section
\ref{Sec-sysmodel}, we describe the system model and the problem
formulation. In Section \ref{Sec-WBoth}, we present the results
obtained for the case when SIC decoders are available at both the BS
and the WiFi AP. In Section \ref{Sec-WOBoth}, we present the results
obtained for the case when SIC decoders are not available at both
the BS and the WiFi AP, and the results for the case when the SIC
decoder is available at the BS side are given in Section
\ref{Sec-WOneSide}. Then, in Section \ref{Sec-extension}, we show
that SIC decoders are beneficial for the cellular operator. We also present a high-efficiency distributed data offloading scheme for the case when SIC decoders are available at both the BS and the WiFi AP. Simulation results are given in
Section \ref{NumericalResults}.  Section \ref{Sec-conclusion} concludes the paper.

\section{System Model}\label{Sec-sysmodel}
\begin{figure}[htb]
        \centering
        \includegraphics*[width=8cm]{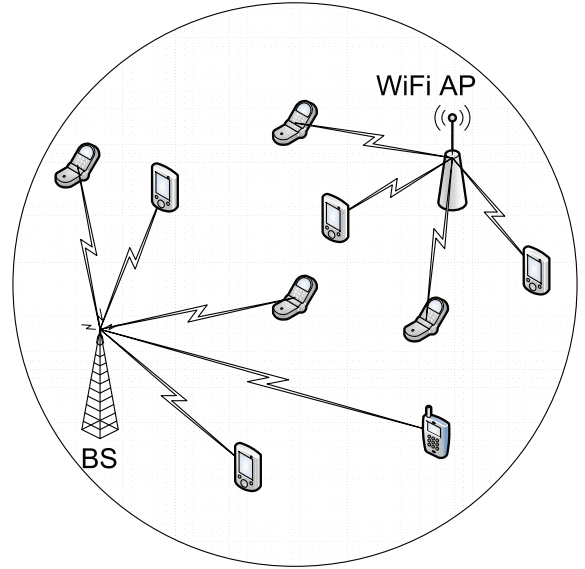}%\vspace{-3mm} %*[width=8cm]
        \caption{System Model}%\vspace{-3mm}
        \label{Fig-Systemmodel}
\end{figure}
In this paper, as shown in Fig. \ref{Fig-Systemmodel}, we consider a
cellular network with $N$ users served by a base station (BS). We
assume that there is a third-party WiFi access point (AP) within the
coverage area of the BS. The WiFi AP and the BS use orthogonal
frequencies. Thus, there is no inter-network interference between
WiFi and cellular network. To maximize the network throughput and
improve the overall network performance, the cellular operator may
direct some of its users to be served by the WiFi AP. Since the WiFi
AP belongs to a third-party operator, data offloading through AP is
thus not for free. The cellular operator has to reward the AP
operator an incentive while guaranteeing an optimized utility.

In this paper, we focus on the uplink scenario. We assume that all
the users adopt fixed power transmission, i.e., $P_i$ for user $i$.
For the convenience of analysis, we assume that $P_i=P, \forall i$.
We also assume the users are uniformly distributed in the coverage
area. The channel power gain between user $i$ and the BS is denoted
by $g_{i,B}$, and that between user $i$ and the WiFi AP is denoted
by $g_{i,A}$. Unless otherwise specified, we assume that $g_{i,B}$'s
and $g_{i,A}$'s are strictly positive, mutually independent, and
have continuous probability distribution function (pdf). The power
of the additive Gaussian noises at the BS and the AP are denoted by
$\sigma^2_{B}$ and $\sigma^2_{A}$, respectively. We also assume that
all the channel state information (CSI) and users' transmit power
are known at the BS. Now, we define $x_i \in \{0, 1\}$ and $y_i \in
\{0, 1\}$ as two indicator functions to indicate user $i$'s
connection to BS and AP, respectively. If user $i$ connects to BS,
$x_i=1$; otherwise, $x_i=0 $. Similarly, if user $i$ connects to AP,
$y_i=1$; otherwise, $y_i=0 $. Besides, at any time, user $i$ is only
allowed to connect to either BS or AP, but not to both of them
simultaneously, i.e., $x_i+y_i\le 1, \forall i$.

In this paper, we assume that the cellular operator charges its
users at $\lambda$ per nat of data usage, and it pays the
third-party WiFi operator at $\mu$ per nat of data usage over the
AP. For convenience, throughout the paper, we use the natural
logarithm. Hence, the data is measured in nats rather than in bits.
Then, the utility function of the operator is defined as
\begin{align}\label{eq-utilityobj}
U(\boldsymbol{x},\boldsymbol{y})&=\lambda
R_B(\boldsymbol{x})+(\lambda-\mu)R_A(\boldsymbol{y}),
\end{align}
where $R_B(\boldsymbol{x})$ is the sum-rate at the BS, and
$R_A(\boldsymbol{y})$ is the sum-rate at the WiFi AP. The exact form
of the sum-rate depends on whether the SIC decoder is available.  As
implied by the name, in a receiver with a SIC decoder, users'
signals are extracted from the composite received signal
successively, rather than in parallel. The SIC decoder is able to
remove the interference of the most recently decoded user from the
current composite received signal by subtracting it out. According
to \cite{ITBook},  if a SIC decoder is available at the BS, the
sum-rate at the BS can be written as
$R_B^w(\boldsymbol{x})=\ln\left(1+\sum_{i=1}^N
\frac{g_{i,B}P}{\sigma^2_{B}}x_i\right)$; on the other hand, if the
SIC decoder is  not available at the BS,
$R_B^o(\boldsymbol{x})=\sum_{i=1}^N
\kern-0.5mm\ln\kern-0.5mm\left(\kern-0.5mm1\kern-0.5mm+\kern-0.5mm
\frac{x_i g_{i,B}P}{\sum_{j=1,j\neq i}^N x_j g_{j,B}P+\sigma_B^2}
\kern-0.5mm\right)$. Similarly, at the WiFi AP, we have
$R_A^w(\boldsymbol{y})=\ln\left(1+\sum_{i=1}^N
\frac{g_{i,A}P}{\sigma^2_{A}}y_i\right)$ and
$R_A^o(\boldsymbol{y})=\sum_{i=1}^N
\kern-0.5mm\ln\kern-0.5mm\left(\kern-0.5mm1\kern-0.5mm+\kern-0.5mm
\frac{y_i g_{i,A}P}{\sum_{j=1,j\neq i}^N y_j g_{j,A}P+\sigma_A^2}
\kern-0.5mm\right)$, with or without SIC decoder.

Depending on whether SIC decoder is available at the BS/AP, we have
the following four possible utility functions
%\begin{itemize}
%  \item With SIC Decoder at both the BS and the AP
%  \item Without SIC Decoder at both the BS and the AP
%  \item With SIC Decoder at both the BS and the AP
%  \item With SIC Decoder at both the BS and the AP
%\end{itemize}
\begin{align}
U^{ww}(\boldsymbol{x},\boldsymbol{y})&=\lambda
R_B^w(\boldsymbol{x})+(\lambda-\mu)R_A^w(\boldsymbol{y}),\\
U^{oo}(\boldsymbol{x},\boldsymbol{y})&=\lambda
R_B^o(\boldsymbol{x})+(\lambda-\mu)R_A^o(\boldsymbol{y}),\\
U^{wo}(\boldsymbol{x},\boldsymbol{y})&=\lambda
R_B^w(\boldsymbol{x})+(\lambda-\mu)R_A^o(\boldsymbol{y}),\\
U^{ow}(\boldsymbol{x},\boldsymbol{y})&=\lambda
R_B^o(\boldsymbol{x})+(\lambda-\mu)R_A^w(\boldsymbol{y}).
\end{align}
In the rest of the paper, we study the optimal data offloading
schemes for the above four cases.

% and assume that
%both the BS and the AP adopt successive cancellation decoding

\section{With SIC Decoders at Both Sides}\label{Sec-WBoth}
In this Section, we investigate the case that both the BS and the
WiFi AP are equipped with a SIC decoder. Thus, the utility
maximization problem of the cellular operator can be formulated as

\underline{\textbf{Problem 3.1}}:
\begin{align} \max_{\{x_i,y_i,\forall i\}} ~&\lambda
\ln\kern-0.5mm\left(\kern-0.5mm1\kern-0.5mm+\kern-0.5mm\sum_{i=1}^N
S_{i,B}x_i\kern-0.5mm\right)\kern-1mm+\kern-0.5mm(\lambda\kern-0.5mm-\kern-0.5mm\mu)\ln\kern-0.5mm\left(\kern-0.5mm1\kern-0.5mm+\kern-0.5mm\sum_{i=1}^N
S_{i,A}y_i\kern-0.5mm\right),\label{Eq-obj}\\
\mbox{s.t.}~~&  x_i \in \{0, 1\}, ~\forall i,\label{Eq-Ineq1}\\
& y_i \in \{0, 1\}, ~\forall i,\label{Eq-Ineq2}\\
& x_i+y_i\le 1, ~\forall i, \label{Eq-Ineq3}
\end{align}
where $S_{i,B}\triangleq \frac{g_{i,B}P}{\sigma^2_{B}}$ and $S_{i,A}
\triangleq \frac{g_{i,A}P}{\sigma^2_{A}}$.

It is observed from this problem formulation that the third-party
operator's pricing strategy $\mu$ has a great influence on the
optimal solution of the above problem. When $\mu$ is larger than
$\lambda$, the cellular operator will not assign any user to the AP.
This is rigorously proved by the following proposition.

\underline{\textbf{Proposition 3.1}}: When $\lambda\le \mu$, the
optimal solution of Problem 3.1 is
$\boldsymbol{x}^*=\boldsymbol{1}_N,
\boldsymbol{y}^*=\boldsymbol{0}_N$, where $\boldsymbol{1}_N$ and
$\boldsymbol{0}_N$ denote the N-dimension all-one vector and
all-zero vector, respectively.

\emph{Proof:} To prove $\boldsymbol{x}^*=\boldsymbol{1}_N$ and
$\boldsymbol{y}^*=\boldsymbol{0}_N$ is the optimal solution of
Problem 3.1, we have to show that $f(\boldsymbol{x}^*,
\boldsymbol{y}^*)$ is larger than $f(\boldsymbol{x},
\boldsymbol{y})$, where $f(\boldsymbol{x}, \boldsymbol{y})$ denotes
the objective function of Problem 3.1 and $(\boldsymbol{x},
\boldsymbol{y})$ is any feasible solution of Problem 3.1. Suppose
$(\tilde{\boldsymbol{x}}, \tilde{\boldsymbol{y}})$ is a feasible
solution of Problem 3.1, then it follows that
\begin{align}
f(\tilde{\boldsymbol{x}}, \tilde{\boldsymbol{y}})&=\lambda R^w_B(\tilde{\boldsymbol{x}})+(\lambda-\mu)R^w_A(\tilde{\boldsymbol{y}})\nonumber\\
 &\stackrel{a}{\le} \lambda R^w_B(\tilde{\boldsymbol{x}})+(\lambda-\mu)R^w_A(\boldsymbol{0}_N)\nonumber\\
 &\stackrel{b}{\le} \lambda
 R^w_B(\boldsymbol{1}_N)+(\lambda-\mu)R^w_A(\boldsymbol{0}_N),
\end{align}
where ``a'' follows from the fact that $\lambda-\mu \le 0$ and
$R^w_A(\boldsymbol{y})$ is always nonnegative, and ``b'' follows
from the fact that $R^w_B(\boldsymbol{x})$ is an increasing function
of $\boldsymbol{x}$, and thus the equality holds only when
$\boldsymbol{x}^*=\boldsymbol{1}_N$. \hfill $\square$

Proposition 3.1 indicates that the cellular operator will not
offload any mobile data to the WiFi AP if the third-party operator
charges at a price higher than its revenue, i.e., $\mu\ge \lambda$.
On the other hand, from the third-party operator's perspective, if
the cellular operator does not offload mobile data through its WiFi
AP, it will earn nothing, which is a lose-lose situation. Thus, a
reasonable third-party operator will charge a price lower than
$\lambda$, which is the scenario we consider in the following
studies, i.e., $\mu<\lambda$.

\underline{\textbf{Proposition 3.2}}: The optimal solution of
Problem 3.1 is obtained when \eqref{Eq-Ineq3} holds with equality
for arbitrary $i$.

\emph{Proof:} This can be proved by contradiction. Suppose
$(\boldsymbol{x}^*,\boldsymbol{y}^*)$ is the optimal solution of
Problem 3.1, and it has an element $(x_k^*,y_k^*)$ satisfying
$x^*_k+y^*_k<1$. Then, from \eqref{Eq-Ineq1} and \eqref{Eq-Ineq2},
it follows that $x^*_k=0, y^*_k=0$. Now, we show that we can always
find a feasible solution
$(\tilde{\boldsymbol{x}},\tilde{\boldsymbol{y}})$ with its elements
satisfying $\tilde{x}^*_i+\tilde{y}^*_i=1, \forall i$ with a higher
value of \eqref{Eq-obj}. We let
$\tilde{\boldsymbol{x}}_{-k}=\boldsymbol{x}^*_{-k}$,
$\tilde{\boldsymbol{y}}_{-k}=\boldsymbol{y}^*_{-k}$,
\textcolor[rgb]{0.00,0.00,0.00}{where the minus sign before the
letter $k$ in the subscript of a vector refers to all the elements
of the vector except the $k$th element}. Then, since the logarithm
function is an increasing function, it is clear that if we set
$\tilde{x}^*_k=1, \tilde{y}^*_k=0$ or $\tilde{x}^*_k=0,
\tilde{y}^*_k=1$ will result in a higher value of \eqref{Eq-obj}
than that resulted by $x^*_k=0, y^*_k=0$. This contradicts with our
presumption. Proposition 3.2 is thus proved. \hfill $\square$

With the results given in Proposition 3.2, we can reduce the
complexity of Problem 3.1 by setting $y_i=1-x_i$.  Problem 3.1 can
be converted to the following problem.

\underline{\textbf{Problem 3.2}}:
\begin{align}
\max_{x_i,\forall i} ~&\lambda
\ln\kern-0.5mm\left(\kern-0.5mm1\kern-1mm+\kern-1mm\sum_{i=1}^N
S_{i,B}x_i\kern-0.5mm\right)\kern-1mm+\kern-0.5mm(\lambda\kern-0.5mm-\kern-0.5mm\mu)\ln\kern-0.5mm\left(\kern-0.5mm1\kern-0.5mm+\kern-0.5mm\sum_{i=1}^N
S_{i,A}(1\kern-0.5mm-\kern-0.5mmx_i)\kern-0.5mm\right),\\
\mbox{s.t.}~~& x_i \in \{0, 1\}, ~\forall i.
\end{align}

This is a nonlinear integer programming problem. When the number of
users is small, it can be solved by exhaustive search. However, when
the number of users is large, exhaustive search is not applicable
due to the high complexity. In this paper, we solve Problem 3.2 by
solving its relaxation problem, and rigorously prove that the gap
between the relaxation problem and Problem 3.2 is negligible when
the number of the users is large.

The \textbf{relaxation problem} of Problem 3.2 is given as follows:

\underline{\textbf{Problem 3.3}}:
\begin{align}
\max_{x_i, \forall i} ~&\lambda
\ln\kern-0.5mm\left(\kern-0.5mm1\kern-1mm+\kern-1mm\sum_{i=1}^N
S_{i,B}x_i\kern-0.5mm\right)\kern-1mm+\kern-0.5mm(\lambda\kern-0.5mm-\kern-0.5mm\mu)\ln\kern-0.5mm\left(\kern-0.5mm1\kern-0.5mm+\kern-0.5mm\sum_{i=1}^N
S_{i,A}(1\kern-0.5mm-\kern-0.5mmx_i)\kern-0.5mm\right),\\
\mbox{s.t.}~~& 0\le x_i \le 1, ~\forall i.
\end{align}

Problem 3.3 is  a convex optimization problem. To show its
convexity, we only need to show that the objective function is
convex or concave since all the constraints are linear. Denote the
objective function  of the relaxation problem as $f_r$, then $f_r$
is convex/concave if its Hessian is positive/negative semidefinite.
Denote the Hessian of $f_r$ as $\boldsymbol{H}$, we show that
$\boldsymbol{H}$ is negative semidefinite by the following
proposition.

\underline{\textbf{Proposition 3.3}}: The Hessian $\boldsymbol{H}$
is negative semidefinite.

\emph{Proof:} The Hessian of $f$ can be written as
\begin{align}
\boldsymbol{H}=\left(
    \begin{array}{ccc}
      \frac{\partial^2 f}{\partial x_1^2} & \cdots & \frac{\partial^2 f}{\partial x_1 \partial x_N} \\
      \vdots& \ddots & \vdots \\
     \frac{\partial^2 f}{\partial x_N \partial x_1} & \cdots & \frac{\partial^2 f}{\partial x_N^2} \\
    \end{array}
  \right),
\end{align}
where the diagonal elements and off-diagonal elements can be
obtained as $ \frac{\partial^2 f}{\partial
x_k^2}\kern-0.5mm=\kern-0.5mm-\frac{\lambda
S_{k,B}^2}{\left(1+\kern-0.5mm\sum_{i=1}^N
S_{i,B}x_i\right)^2}\kern-0.5mm-\kern-0.5mm\frac{(\lambda-\mu)
S_{k,A}^2}{\left(1+\kern-0.5mm\sum_{i=1}^N
S_{i,A}(1\kern-0.5mm-x_i)\right)^2}, $ and $ \frac{\partial^2
f}{\partial x_k \partial x_j}\kern-0.5mm=\kern-0.5mm-\frac{\lambda
S_{k,B}S_{j,B}}{\left(1+\sum_{i=1}^N\kern-0.5mm
S_{i,B}x_i\right)^2}-\kern-0.5mm\frac{(\lambda-\mu)
S_{k,A}S_{j,A}}{\left(1+\sum_{i=1}^N\kern-0.5mm
S_{i,A}(1\kern-0.5mm-x_i)\right)^2}. $
%\begin{align}
%\frac{\partial^2 f}{\partial
%x_k^2}&\kern-0.5mm=\kern-0.5mm-\frac{\lambda
%S_{k,B}^2}{\left(1\kern-0.5mm+\kern-0.5mm\sum_{i=1}^N
%S_{i,B}x_i\right)^2}\kern-0.5mm-\kern-0.5mm\frac{(\lambda\kern-0.5mm-\kern-0.5mm\mu)
%S_{k,A}^2}{\left(1\kern-0.5mm+\kern-0.5mm\sum_{i=1}^N
%S_{i,A}(1\kern-0.5mm-\kern-0.5mmx_i)\right)^2},
%\end{align}
%\begin{align}
%\frac{\partial^2 f}{\partial x_k \partial
%x_j}&\kern-0.5mm=\kern-0.5mm-\frac{\lambda
%S_{k,B}S_{j,B}}{\left(1\kern-1mm+\kern-1mm\sum_{i=1}^N\kern-0.5mm
%S_{i,B}x_i\right)^2}-\kern-0.5mm\frac{(\lambda\kern-0.5mm-\kern-0.5mm\mu)
%S_{k,A}S_{j,A}}{\left(1\kern-1mm+\kern-1mm\sum_{i=1}^N\kern-0.5mm
%S_{i,A}(1\kern-0.5mm-\kern-0.5mmx_i)\right)^2}.
%\end{align}

It is observed $\boldsymbol{H}$ can be rewritten as
\begin{align}
\boldsymbol{H}\kern-0.5mm=\kern-0.5mm-\frac{\lambda}{\left(1\kern-0.5mm+\kern-0.5mm\sum_{i=1}^N\kern-0.5mm
S_{i,B}x_i\right)^2}
\boldsymbol{B}\kern-0.5mm-\kern-0.5mm\frac{(\lambda\kern-0.5mm-\kern-0.5mm\mu)}{\left(1\kern-0.5mm+\kern-0.5mm\sum_{i=1}^N\kern-0.5mm
S_{i,A}(1\kern-0.5mm-\kern-0.5mmx_i)\right)^2}\boldsymbol{A},
\end{align}
where matrices $\boldsymbol{B}$ and $\boldsymbol{A}$ have the same
structure as the following matrix $\boldsymbol{X}$
\begin{align}
\boldsymbol{X}=\left(
    \begin{array}{ccc}
      S_{1,X}^2 & \cdots & S_{1,X}S_{N,X}\\
      \vdots& \ddots & \vdots \\
      S_{N,X}S_{1,X} & \cdots & S_{N,X}^2\\
    \end{array}
  \right).
\end{align}

It can be shown that for any vector $\boldsymbol{c}=[c_1~ \cdots~
c_N]^T$, $\boldsymbol{c}^T \boldsymbol{X} \boldsymbol{c}$ can be
obtained as
\begin{align}\boldsymbol{c}^T \boldsymbol{X} \boldsymbol{c}=\left(c_1S_{1,X}+\cdots+c_N S_{N,X}\right)^2 \ge
0.\end{align} Thus, it is clear that both $\boldsymbol{B}$ and
$\boldsymbol{A}$ are positive semidefinite. Then, since both
$\lambda$ and $\lambda-\mu$ are non-negative, it is easy to see that
$\boldsymbol{H}$ is negative semidefinite. Therefore, the objective
function is strictly
concave. \hfill $\square$

Problem 3.3 is shown to be convex, and it can be easily verified
that Slater's condition holds for this problem. Thus, the duality
gap between Problem 3.3 and its dual problem is zero, and solving
its dual problem is equivalent to solving the original problem.

Now, we consider its dual problem. The Lagrangian of Problem 3.3 is
\begin{align}
L\left(\boldsymbol{x}, \boldsymbol{\alpha},
\boldsymbol{\beta}\right)=(\lambda-\kern-0.5mm\mu)\ln\kern-0.5mm\left(\kern-0.5mm1\kern-0.5mm+\kern-0.5mm\sum_{i=1}^N
S_{i,A}(1-x_i)\kern-0.5mm\right)\nonumber\\+\lambda
\ln\kern-0.5mm\left(\kern-0.5mm1\kern-0.5mm+\kern-0.5mm\sum_{i=1}^N
S_{i,B}x_i\kern-0.5mm\right)-\sum_{i=1}^{N}\alpha_i
(x_i-1)+\sum_{i=1}^{N}\beta_i x_i,
\end{align}
where $\boldsymbol{\alpha}=[\alpha_1 ~\cdots~ \alpha_N]^T$ and
$\boldsymbol{\beta}=[\beta_1 ~\cdots~ \beta_N]^T$ are the
nonnegative dual variables associated with the constraints.

The dual function is $q(\boldsymbol{\alpha},
\boldsymbol{\beta})=\max_{\boldsymbol{x}} L(\boldsymbol{x},
\boldsymbol{\alpha}, \boldsymbol{\beta})$. The Lagrange dual problem
is then given by $\min_{\boldsymbol{\alpha} \succcurlyeq 0,
\boldsymbol{\beta} \succcurlyeq 0} q(\boldsymbol{\alpha},
\boldsymbol{\beta})$. Therefore, the optimal solution needs to
satisfy the following Karush-Kuhn-Tucker (KKT) conditions
\cite{kangTWC}:
\begin{align}
\alpha_k (x_k^*-1)&=0,\label{eq-KKT02}\\
\beta_k x_k^*&=0,\label{eq-KKT03}\\
0\le x_k^* &\le 1,\\
\alpha_k^* \ge0, ~ \beta_k^* &\ge0,\\
\frac{\partial L\left(\boldsymbol{x}, \boldsymbol{\alpha},
\boldsymbol{\beta}\right)}{\partial x_k^*}=-\frac{(\lambda-\mu)
S_{k,A}}{1\kern-0.5mm+\kern-0.5mm\sum_{i=1}^N
S_{i,A}(1-x_i^*)}~~~~~~&\nonumber\\+\frac{\lambda
S_{k,B}}{1\kern-0.5mm+\kern-0.5mm\sum_{i=1}^N
S_{i,B}x_i^*}-\alpha_k+\beta_k&=0,\label{eq-KKT01}
\end{align}
Due to the complexity of the problem, solving the above KKT
conditions  will not render us a closed-form solution. However, from
these KKT conditions, we are able to gain some significant features
of the optimal solution.
%we are not able to
%obtain the closed-form solution by solving these KKT conditions.

\underline{\textbf{Theorem 3.1}}: The optimal solution of the
relaxation problem has at most one user indexed by $k$ ($k\in
\left\{1,2,\cdots,N\right\}$), with a fractional $x_k$ satisfying
$0<x_k<1$.

\emph{Proof:} This proposition can be proved by contradiction. Suppose that there
are two arbitrary users denoted by $m$ and $n$ having fractional
$x_m$ and $x_n$, respectively, i.e., $0<x_m<1$ and $0<x_n<1$. From
\eqref{eq-KKT02} and \eqref{eq-KKT03}, it follows that $\alpha_m=0$,
$\alpha_n=0$, $\beta_m=0$, and $\beta_n=0$. Then, applying these
facts to \eqref{eq-KKT01}, it follows that
\begin{align}
\frac{\lambda S_{m,B}}{1\kern-0.5mm+\kern-0.5mm\sum_{i=1}^N
S_{i,B}x_i^*}-\frac{(\lambda-\mu)
S_{m,A}}{1\kern-0.5mm+\kern-0.5mm\sum_{i=1}^N
S_{i,A}(1-x_i^*)}=0,\\
\frac{\lambda S_{n,B}}{1\kern-0.5mm+\kern-0.5mm\sum_{i=1}^N
S_{i,B}x_i^*}-\frac{(\lambda-\mu)
S_{n,A}}{1\kern-0.5mm+\kern-0.5mm\sum_{i=1}^N S_{i,A}(1-x_i^*)}=0.
\end{align}
Then, for these two users, the following equality must hold
%\begin{align}
%\frac{ S_{m,B}}{S_{m,A}}=\frac{S_{n,B}}{S_{n,A}},
%\end{align}
\begin{align}
\frac{
S_{m,B}}{S_{m,A}}=\frac{S_{n,B}}{S_{n,A}}=\frac{(\lambda-\mu)}{\lambda}\frac{1\kern-0.5mm+\kern-0.5mm\sum_{i=1}^N
S_{i,B}x_i^*}{1\kern-0.5mm+\kern-0.5mm\sum_{i=1}^N
S_{i,A}(1-x_i^*)}.
\end{align}
%\begin{align}
%\frac{
%S_{m,B}-S_{n,B}}{S_{m,A}-S_{n,A}}=\frac{(\lambda-\mu)}{\lambda}\frac{1\kern-0.5mm+\kern-0.5mm\sum_{i=1}^N
%S_{i,B}x_i}{1\kern-0.5mm+\kern-0.5mm\sum_{i=1}^N S_{i,A}(1-x_i)}.
%\end{align}
It is easy to observe that $\frac{
S_{m,B}}{S_{m,A}}=\frac{S_{n,B}}{S_{n,A}}$ is equivalent to $ \frac{
g_{m,B}}{g_{m,A}}=\frac{g_{n,B}}{g_{n,A}}$, However, it can be
verified that the equality $ \frac{
g_{m,B}}{g_{m,A}}=\frac{g_{n,B}}{g_{n,A}}$ is satisfied with a zero
probability since the channel power gains are mutually independent
and have continuous pdf. This result contradicts with our
presumption. Thus, it is concluded that there is at most one user
with a fractional $x_k$, i.e., $0<x_k<1$. Theorem 3.1 is thus
proved.\hfill $\square$

From Theorem 3.1, it is observed that there is at most one user with
a fractional indicator for the optimal solution of Problem 3.3. This
indicates that the optimal solution of Problem 3.3 is either equal
to or just one-user away from that of Problem 3.2. Thus, the
following scheme is proposed to find the optimal solution of Problem
3.2.

\begin{table}[!h]
\tabcolsep 0pt \caption{}  \vspace*{-10pt}
\begin{center}
\def\temptablewidth{0.48\textwidth}
{\rule{\temptablewidth}{1pt}}
\begin{tabular*}{\temptablewidth}{@{\extracolsep{\fill}}l}
Proposed Centralized Data Offloading Scheme for Problem 3.1  \\
\hline 1). Solve Problem 3.3 by standard convex
optimization algorithms, such as\\ interior-point method\cite{ConvexOpt},  or existing solvers such as CVX\cite{CVX}.\\
2). Convert the obtained solution into a feasible solution of
Problem 3.2 by\\ rounding the fractional indicator function to its
nearest integer ($0$ or $1$).
\end{tabular*}
{\rule{\temptablewidth}{1pt}}
\end{center}
\end{table}
%\vspace*{-8pt}

In general, the above algorithm provides a sub-optimal solution to
Problem 3.2. However, due to the special feature presented in
Theorem 3.1, we are able to prove that the proposed solution given
in the Table I is near-optimal when the number of users is large.

\underline{\textbf{Theorem 3.2}}: The gap between the optimal
solution of Problem 3.2 and the proposed solution given in Table I
is negligible when the number of users is large.

\emph{Proof:} For the convenience of exposition, we denote the maximum
values of Problem 3.2 attained at the optimal solution and at the
proposed solution given in Table I as $f_o^*$ and $f_s^*$,
respectively. Since the solution given in Table I is also a feasible
solution of Problem 3.2. Thus, it follows that
\begin{align}
f_s^* \le f_o^*.\end{align}

On the other hand, it is clear that the maximum value of Problem 3.2
is upper bounded by its relaxation problem. Thus, if we denote the
maximum values of the relaxation problem attained at the optimal
solution as $f_r^*$, it follows that
\begin{align}f_o^* \le f_r^*.\end{align}

Combining the above facts together, we have
\begin{align}f_s^* \le
f_o^* \le f_r^*.\end{align} Thus, if we are able to show that the
gap between $f_s^*$ and $f_r^*$ is negligible when the number of
users is large, it is clear that the gap between $f_s^*$ and $f_o^*$
will also be negligible when the number of users is large.

Now, we show that the gap between $f_s^*$ and $f_r^*$ is negligible
when the number of users is large. Suppose $\boldsymbol{x}^*$ is the
optimal solution of the relaxation problem, and user $k$ is the user
with a fractional indicator function $x_k^*$. Then, the value of
$f_r^*$ is $\lambda
\ln\kern-0.5mm\left(\kern-0.5mm1\kern-0.5mm+\kern-0.5mm\sum_{i=1}^N
S_{i,B}x_i^*\kern-0.5mm\right)\kern-1mm+(\lambda-\kern-0.5mm\mu)\ln\kern-0.5mm\left(\kern-0.5mm1\kern-0.5mm+\kern-0.5mm\sum_{i=1}^N
S_{i,A}(1-x_i^*)\kern-0.5mm\right),$
%\begin{align}\label{Eq-reobj1}
%&\lambda
%\ln\kern-0.5mm\left(\kern-0.5mm1\kern-0.5mm+\kern-0.5mm\sum_{i=1}^N
%S_{i,B}x_i^*\kern-0.5mm\right)\kern-1mm+(\lambda-\kern-0.5mm\mu)\ln\kern-0.5mm\left(\kern-0.5mm1\kern-0.5mm+\kern-0.5mm\sum_{i=1}^N
%S_{i,A}(1-x_i^*)\kern-0.5mm\right).
%\end{align}
while the value of $f_s^*$ is obtained by either setting $x_k=0$
when $x_k<0.5$ or setting $x_k=1$ when $x_k\ge 0.5$. Obviously, it
follows that $f_s^*>\tilde{f_s}^*$, where $\tilde{f_s}^*\triangleq
\lambda \ln\left(1+\sum_{i=1,i\neq k}^N
S_{i,B}x_i^*\right)+(\lambda-\mu)\ln\left(1+\sum_{i=1,i\neq k}^N
S_{i,A}(1-x_i^*)\right),$ which corresponds to the scenario that
user $k$ connects to neither the BS nor the AP.
%\begin{align}\label{Eq-reobj2}
%&\lambda
%\ln\kern-0.5mm\left(\kern-0.5mm1\kern-0.5mm+\kern-2mm\sum_{i=1,i\neq
%k}^N\kern-1.5mm
%S_{i,B}x_i^*\kern-0.5mm\right)\kern-1mm+(\lambda\kern-0.5mm-\kern-0.5mm\mu)\ln\kern-0.5mm\left(\kern-0.5mm1\kern-0.5mm+\kern-1.5mm\sum_{i=1,i\neq
%k}^N\kern-1.5mm S_{i,A}(1-x_i^*)\kern-0.5mm\right).
%\end{align}

Then, the gap $\Delta$ between $f_s^*$ and $f_r^*$ satisfies
\begin{align}
\Delta&< f_r^*-\tilde{f_s}^*= \lambda
\ln\kern-0.5mm\left(1+\frac{S_{k,B}x_k^*\kern-0.5mm}{\kern-0.5mm1\kern-0.5mm+\kern-0.5mm\sum_{i=1,i\neq
k}^N
S_{i,B}x_i^*\kern-0.5mm}\right)\nonumber\\&+(\lambda-\kern-0.5mm\mu)\ln\kern-0.5mm\left(1+\frac{S_{k,A}(1\kern-0.5mm-\kern-0.5mmx_k^*)}{\kern-0.5mm1\kern-0.5mm+\kern-0.5mm\sum_{i=1,i\neq
k}^N S_{i,A}(1-x_i^*)\kern-0.5mm}\right).
\end{align}
%\begin{align}
%&\lambda
%\ln\kern-0.5mm\left(\frac{\kern-0.5mm1\kern-0.5mm+\kern-0.5mm\sum_{i=1,i\neq
%k}^N
%S_{i,B}x_i^*+S_{k,B}x_k^*\kern-0.5mm}{\kern-0.5mm1\kern-0.5mm+\kern-0.5mm\sum_{i=1,i\neq
%k}^N
%S_{i,B}x_i^*\kern-0.5mm}\right)\kern-1mm+(\lambda-\kern-0.5mm\mu)\ln\kern-0.5mm\left(\frac{\kern-0.5mm1\kern-0.5mm+\kern-0.5mm\sum_{i=1,i\neq
%k}^N
%S_{i,A}(1-x_i^*)\kern-0.5mm+S_{k,A}(1\kern-0.5mm-\kern-0.5mmx_k^*)}{\kern-0.5mm1\kern-0.5mm+\kern-0.5mm\sum_{i=1,i\neq
%k}^N S_{i,A}(1-x_i^*)\kern-0.5mm}\right)\nonumber\\
%=&\lambda
%\ln\kern-0.5mm\left(1+\frac{S_{k,B}x_k^*\kern-0.5mm}{\kern-0.5mm1\kern-0.5mm+\kern-0.5mm\sum_{i=1,i\neq
%k}^N
%S_{i,B}x_i^*\kern-0.5mm}\right)\kern-1mm+(\lambda-\kern-0.5mm\mu)\ln\kern-0.5mm\left(1+\frac{S_{k,A}(1\kern-0.5mm-\kern-0.5mmx_k^*)}{\kern-0.5mm1\kern-0.5mm+\kern-0.5mm\sum_{i=1,i\neq
%k}^N S_{i,A}(1-x_i^*)\kern-0.5mm}\right).
%\end{align}
Since the users are uniformly distributed in the area, thus when the
number of users is large, it is inferred that the denominators of
the above equation will be very large. Consequently, the value of
$\Delta$ is close to zero. Theorem 3.2 is thus proved. \hfill $\square$

\section{Without SIC Decoders at Both Sides} \label{Sec-WOBoth}
In this section, we consider the scenario that neither the BS nor
the WiFi AP implements the SIC decoder. The utility maximization
problem of the cellular operator for this case can be formulated as

\underline{\textbf{Problem 4.1}}:
\begin{align}
\max_{\{x_i,y_i,\forall i\}}~~&\lambda \sum_{i=1}^N
\kern-0.5mm\ln\kern-0.5mm\left(\kern-0.5mm1\kern-0.5mm+\kern-0.5mm
\frac{x_i g_{i,B}P}{\sum_{j=1,j\neq i}^N x_j g_{j,B}P+\sigma_B^2}
\kern-0.5mm\right)\kern-1mm\nonumber\\+(\lambda-\kern-0.5mm\mu)&\sum_{i=1}^N
\kern-0.5mm\ln\kern-0.5mm\left(\kern-0.5mm1\kern-0.5mm+\kern-0.5mm
\frac{y_i g_{i,A}P}{\sum_{j=1,j\neq i}^N y_j g_{j,A}P+\sigma_A^2}
\kern-0.5mm\right),\label{Eq-WOSICobj}\\
\mbox{s.t.}~~~~&  x_i \in \{0, 1\}, ~\forall i,\label{Eq-WOSICIneq1}\\
& y_i \in \{0, 1\}, ~\forall i,\label{Eq-WOSICIneq2}\\
& x_i+y_i\le 1, ~\forall i. \label{Eq-WOSICIneq3}
\end{align}
Problem 4.1 is a nonlinear integer programming problem which is
difficult to solve directly due to its high complexity. Besides, it
is not difficult to verify that the relaxation problem of Problem
4.1 is non-convex. Thus, we are not able to solve Problem 4.1 in the
same way as Problem 3.1. To solve Problem 4.1, we first consider the
following two subproblems.

\underline{\textbf{Subproblem 4.1a}}:
\begin{align}
\max_{\{x_i,\forall i\}}~~&\lambda \sum_{i=1}^N
\kern-0.5mm\ln\kern-0.5mm\left(\kern-0.5mm1\kern-0.5mm+\kern-0.5mm
\frac{x_i g_{i,B}P}{\sum_{j=1,j\neq i}^N x_j g_{j,B}P+\sigma_B^2}
\kern-0.5mm\right),\\
\mbox{s.t.}~~~~&  x_i \in \{0, 1\}, ~\forall i.
\end{align}

\underline{\textbf{Subproblem 4.1b}}:
\begin{align}
\max_{\{y_i,\forall i\}}~~&(\lambda-\kern-0.5mm\mu)\sum_{i=1}^N
\kern-0.5mm\ln\kern-0.5mm\left(\kern-0.5mm1\kern-0.5mm+\kern-0.5mm
\frac{y_i g_{i,A}P}{\sum_{j=1,j\neq i}^N y_j g_{j,A}P+\sigma_A^2}
\kern-0.5mm\right),\\
\mbox{s.t.}~~~~ & y_i \in \{0, 1\}, ~\forall i.
\end{align}

Denote the optimal solution of Subproblem 4.1a as $x_i^*,~\forall i
\in \{1,2,\cdots,N\}$ and that of Subproblem 4.1b as $y_i^*,~\forall
i \in \{1,2,\cdots,N\}$. Then, it is clear that if $x_i^*$ and
$y_i^*$ satisfy the constraints \eqref{Eq-WOSICIneq3} for all $i\in
\{1,2,\cdots,N\}$, $x_i^*$ and $y_i^*$ will be the optimal solution
for Problem 4.1. In the following, we will show that when the number
of users is large, Problem 4.1 can be solved by individually solving
Subproblem 4.1a and Subproblem 4.1b. It is seen that Subproblem 4.1a
and Subproblem 4.1b have the same structure.  As a result, the
optimal solutions of these two subproblems should also have the same
structure. In the following, using Subproblem 4.1b as an example, we
present the optimal solution of the two subproblems.

\underline{\textbf{Lemma 4.1}}: i) Sort the users according to their
channel power gains in descending order: $g_{1,A}\ge g_{2,A}\ge
\cdots \ge g_{N,A}$. At an optimal solution, only the first $k^*
(\le N)$ users transmit, and
$k^*=\mbox{argmax}_{k}~\sum_{i=1}^k\ln\left(1+\frac{g_{i,A}P}{\sum_{j=1,j\neq
i}^k g_{j,A}P+\sigma_A^2}\right)$.

ii) Further, if $g_{1,A}\ge \frac{\left(e-1\right)\sigma_A^2}{P}$, $k^*=1$. That is, only the user with the largest channel gain transmits.

\emph{Proof:} To solve Subproblem 4.1b, we first consider its relaxation problem,
which is given as follows.

\underline{\textbf{Problem 4.2}}
\begin{align}
\max_{\{y_i,\forall i\}}~~&\sum_{i=1}^N
\kern-0.5mm\ln\kern-0.5mm\left(\kern-0.5mm1\kern-0.5mm+\kern-0.5mm
\frac{y_i g_{i,A}P}{\sum_{j=1,j\neq i}^N y_j g_{j,A}P+\sigma_A^2}
\kern-0.5mm\right),\\
\mbox{s.t.}~~~~ & 0\le y_i \le 1, ~\forall i.
\end{align}
Let $P_i\triangleq y_{i}P, \forall i$, it is not difficult to
observe that Problem 4.2 can be converted to the following problem,

\underline{\textbf{Problem 4.3}}
\begin{align}
\max_{\{P_i,\forall i\}}~~&\sum_{i=1}^N
\kern-0.5mm\ln\kern-0.5mm\left(\kern-0.5mm1\kern-0.5mm+\kern-0.5mm
\frac{ g_{i,A}P_i}{\sum_{j=1,j\neq i}^N g_{j,A}P_j +\sigma_A^2}
\kern-0.5mm\right),\\
\mbox{s.t.}~~~~ & 0\le P_i \le P, ~\forall i.
\end{align}
This problem is shown to be Schur convex  in \cite{Hanly}. By using
the Schur convex properties, it is shown in \cite{Hanly} that the
optimal power allocation is binary power allocation, i.e., either
$0$ or $P$ for all $i$. This indicates that the optimal solution for
Problem 4.2 is either $0$ or $1$ for all $i$. Thus, it can be
observed that Problem 4.2 is actually equivalent to Subproblem 4.1b.
Then, the results obtained for Problem 4.3 can be directly applied
to Subproblem 4.1b. Based on the results in \cite{Hanly} (Theorem 1
and 4), it is not difficult to obtain the results presented in this
lemma. Details are omitted here for brevity. \hfill $\square$

With the results given in Lemma 4.1, we are now ready for the
following theorem.

\underline{\textbf{Theorem 4.1}}: When the number of users $N$ is
large, with high probability, the optimal solution for Problem 4.1 is
as follows.
\begin{itemize}
  \item Only two users are active in the network: one connects to the BS and the other connects to the WiFi AP.
  \item Denote the index of the users connecting to the BS and the WiFi AP as $m$ and $n$, respectively.  Then,
   user $m$ has the best user-to-BS channel, i.e., $m=\mbox{argmax}_{i}~ g_{i,B}$; and user $n$ has the best user-to-AP channel, i.e., $n=\mbox{argmax}_{i}~g_{i,A}$.
\end{itemize}

\emph{Proof:} Let $m = \mbox{argmax}_{i}~g_{i,B}$ and $n =
\mbox{argmax}_{i}~g_{i,A}$. From Lemma 4.1 Part (ii), (a) if
$g_{n,A}\ge \frac{\left(e-1\right)\sigma_A^2}{P}$, then at the
optimal solution for Subproblem 4.1b, only the user $n$ transmits.
Similarly, (b) if there exists $g_{m,B}\ge
\frac{\left(e-1\right)\sigma_A^2}{P}$, then at the optimal solution
for Subproblem 4.1a, only user $m$ transmits. Finally, (c) if $m\neq
n$, then the optimal solution of Problem 4.1 is the one is given in
Theorem 4.1.

We now show that, when the number of users is large, these three conditions ((a) - (c)) hold with high probability. Define
\begin{itemize}
  \item Event A: There is no user satisfying $g_{i,A}\ge
\frac{\left(e-1\right)\sigma_A^2}{P}$.
  \item Event B: There is no user satisfying
$g_{i,B}\ge \frac{\left(e-1\right)\sigma_B^2}{P}$.
  \item Event C: There exists one
user having the best user-to-BS channel, and simultaneously having
the best user-to-AP channel.
\end{itemize}

Hence, the probability that at least one of the three conditions ((a) - (c)) is violated can be written as
\begin{align}\label{Eq-UnionBound}
\mbox{Prob}\{A\cup B\cup C\} \le
\mbox{Prob}\{A\}+\mbox{Prob}\{B\}+\mbox{Prob}\{C\},
\end{align}
where the inequality results from the well-known union bound.

In the following, we show that $\mbox{Prob}\{A\} \to 0$,
$\mbox{Prob}\{B\}\to 0$, and $\mbox{Prob}\{C\}\to 0$ go to zero as $N \to \infty$. First, we look at
$\mbox{Prob}\{A\}$, which is given by
{\allowdisplaybreaks
\begin{align}
\mbox{Prob}\{A\}&={\rm
Prob}\left\{g_{i,A}<\frac{(e-1)\sigma_A^2}{P}, \forall i\right\}\nonumber\\
&\stackrel{a}{=}\left({\rm
Prob}\left\{g_{A}<\frac{(e-1)\sigma_A^2}{P}\right\}\right)^N\nonumber\\
&=\left(\int_0^{(e-1)\sigma_A^2/P} dF(g_{A})\right)^N,
\end{align}}
where the equality ``a'' results from the fact that the channel
power gains are i.i.d., and $F(g_{A})$ denotes the CDF of the
channel power gain. Since $\int_0^{(e-1)\sigma_A^2/P} dF(g_{A})$ is
strictly less than $1$, $\mbox{Prob}\{A\} \to 0$ as $N \to \infty$.

%\begin{itemize}
%  \item One-user dominant probability $\mathscr{P}$,
%  \item One-user dominant condition $g_{A}>\frac{(e-1)\sigma^2}{P}$,
%\end{itemize}
%\begin{align}
%\mathscr{P}&=1-\left(1-{\rm
%Prob}\left\{g_{A}<\frac{(e-1)\sigma^2}{P}\right\}\right)^N\\
%&=1-\left(1-\int_0^{(e-1)\sigma^2/P} dF(g_{A})\right)^N\end{align}
%It is clear that for given $P$, when $N$ is large, the $\mathscr{P}$
%is close to $1$.

Using the same approach, $\mbox{Prob}\{B\} \to 0$ as $N \to \infty$.

Now, we consider $\mbox{Prob}\{C\}$.
\begin{align}
\mbox{Prob}\{C\}&=\left(\begin{array}{c}
                        N \\
                        1
                      \end{array}
\right){\rm Prob}\left\{\begin{matrix}j=\mbox{argmax}_{i}~ g_{i,A},
\\\mbox{and}
~j=\mbox{argmax}_{i}~ g_{i,B}\end{matrix}\right\}\\
&\stackrel{a}{=}N*{\rm Prob}\left\{j=\mbox{argmax}_{i}~
g_{i,A}\right\}\nonumber\\&~~~~~~*{\rm
Prob}\left\{j=\mbox{argmax}_{i}~ g_{i,B}\right\}\\
&=\frac{1}{N}\label{eq50}
\end{align}
%\begin{align}
%\mbox{Prob}\{C\}&=\left(\begin{array}{c}
%                        N \\
%                        1
%                      \end{array}
%\right){\rm Prob}\left\{j=\mbox{argmax}_{i}~ g_{i,A} ~\mbox{and}
%~j=\mbox{argmax}_{i}~ g_{i,B}\right\}\nonumber\\
%&\stackrel{a}{=}N{\rm Prob}\left\{j=\mbox{argmax}_{i}~
%g_{i,A}\right\}{\rm
%Prob}\left\{j=\mbox{argmax}_{i}~ g_{i,B}\right\}\nonumber\\
%&=\frac{1}{N} \label{eq50}
%\end{align}
where the equality ``a'' results from the fact that the channel
power gains are i.i.d. and have continuous pdf. From~\eqref{eq50},
$\mbox{Prob}\{C\}\to 0$ as $N \to \infty$.
%Then, it is observed
%that $\mathscr{P}_{C}$ is a decreasing function with respect to $N$.
%Thus, it is clear that $\mathscr{P}_C$ goes to $0$ as $N$ goes to
%$+\infty$.

Combining the above results, $1-\mbox{Prob}\{A\cup B\cup
C\}\to 1$ as $N\to \infty$, which completes the proof of Theorem 4.1. \hfill $\square$

\section{With A SIC Decoder at One Side} \label{Sec-WOneSide}
In this section, we consider the scenario that the SIC decoder is
only available at one side. Particularly, we only study the case
that the SIC decoder is only available at the BS. The case that the
SIC decoder is only available at the WiFi AP is a symmetric case,
and thus can be solved in the same way.

\underline{\textbf{Problem 5.1}}:
\begin{align}
\max_{\{x_i,y_i,\forall i\}} ~~&\lambda
\ln\kern-0.5mm\left(\kern-0.5mm1\kern-0.5mm+\kern-0.5mm\sum_{i=1}^N
\frac{x_ig_{i,B}P}{\sigma_B^2}\kern-0.5mm\right)\kern-1mm\nonumber\\+(\lambda-\kern-0.5mm\mu)&\sum_{i=1}^N
\kern-0.5mm\ln\kern-0.5mm\left(\kern-0.5mm1\kern-0.5mm+\kern-0.5mm
\frac{y_i g_{i,A}P}{\sum_{j=1,j\neq i}^N y_j g_{j,A}P+\sigma_A^2}
\kern-0.5mm\right),\label{Eq-WOSICWiFiobj}\\
\mbox{s.t.}~~~~&  x_i \in \{0, 1\}, ~\forall i,\label{Eq-WOSICWiFiIneq1}\\
& y_i \in \{0, 1\}, ~\forall i,\label{Eq-WOSICWiFiIneq2}\\
& x_i+y_i\le 1, ~\forall i. \label{Eq-WOSICWiFiIneq3}
\end{align}

Similar to Problem 4.1, we are not able to solve this problem
directly or by solving its relaxation problem. To solve Problem 5.1,
we first consider the following two subproblems.

\underline{\textbf{Subproblem 5.1a}}:
\begin{align}
\max_{\{x_i,\forall i\}}~~&\lambda
\ln\kern-0.5mm\left(\kern-0.5mm1\kern-0.5mm+\kern-0.5mm\sum_{i=1}^N
\frac{x_ig_{i,B}P}{\sigma_B^2}\kern-0.5mm\right),\\
\mbox{s.t.}~~~~&  x_i \in \{0, 1\}, ~\forall i.
\end{align}

\underline{\textbf{Subproblem 5.1b}}:
\begin{align}
\max_{\{y_i,\forall i\}}~~&(\lambda-\kern-0.5mm\mu)\sum_{i=1}^N
\kern-0.5mm\ln\kern-0.5mm\left(\kern-0.5mm1\kern-0.5mm+\kern-0.5mm
\frac{y_i g_{i,A}P}{\sum_{j=1,j\neq i}^N y_j g_{j,A}P+\sigma_A^2}
\kern-0.5mm\right),\\
\mbox{s.t.}~~~~ & y_i \in \{0, 1\}, ~\forall i.
\end{align}

Denote the optimal solution of Subproblem 5.1a as $x_i^*,~\forall i
\in \{1,2,\cdots,N\}$ and that of Subproblem 5.1b as $y_i^*,~\forall
i \in \{1,2,\cdots,N\}$. Subproblem 5.1a is easy to solve, and the
optimal solution is $x_i^*=1, \forall i$. Subproblem 5.1b is exactly
the same as Subproblem 4.1b, and thus the optimal solution of
Subproblem 5.1b can be obtained from Lemma 4.1. It is obvious that
$x_i^*$ and $y_i^*$ cannot satisfy the constraints
\eqref{Eq-WOSICWiFiIneq3} for all $i\in \{1,2,\cdots,N\}$. Thus,
Problem 5.1 cannot be solved by directly solving Subproblem 5.1a and
Subproblem 5.1b. This makes Problem 5.1 more challenging than
Problem 4.1.

To solve Problem 5.1, we need the following lemma.

\underline{\textbf{Lemma 5.1}}: The optimal solution of Problem 5.1
is obtained when \eqref{Eq-WOSICWiFiIneq3} holds with equality for
all $i$.

%This can be proved by the same method as Proposition 3.2. Thus, the
%proof is omitted here for brevity.
\emph{Proof:} This can be proved by contradiction. Suppose
$(\boldsymbol{x}^*,\boldsymbol{y}^*)$ is the optimal solution of
Problem 5.1, and it has an element $(x_k^*,y_k^*)$ satisfying
$x^*_k+y^*_k<1$. Then, from \eqref{Eq-WOSICWiFiIneq1} and
\eqref{Eq-WOSICWiFiIneq2}, it follows that $x^*_k=0, y^*_k=0$. Now,
we show that we can always find a feasible solution
$(\tilde{\boldsymbol{x}},\tilde{\boldsymbol{y}})$ with its elements
satisfying $x^*_i+y^*_i=1, \forall i$, will result in a higher value
of \eqref{Eq-WOSICWiFiobj}. We let
$\tilde{\boldsymbol{x}}_{-k}=\boldsymbol{x}^*_{-k}$,
$\tilde{\boldsymbol{y}}_{-k}=\boldsymbol{y}^*_{-k}$. Clearly,  if we
set $\tilde{x}^*_k=1, \tilde{y}^*_k=0$ will result in a higher value
of \eqref{Eq-WOSICWiFiobj} than that resulted by $x^*_k=0, y^*_k=0$
since the logarithm function is an increasing function. This
contradicts with our presumption. Lemma 5.1 is thus proved.\hfill $\square$

\begin{figure}[t]
        \centering
        \includegraphics*[width=7cm]{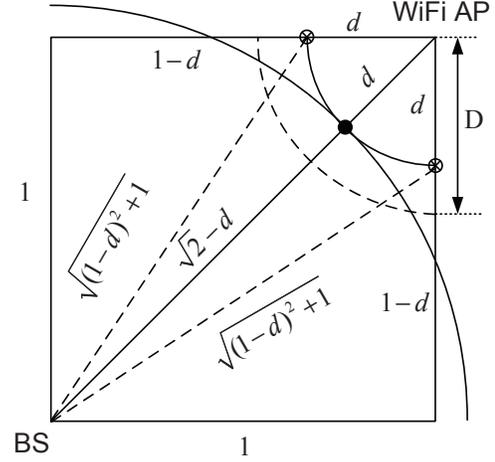}%\vspace{-3mm} %*[width=8cm]
        \caption{Path loss Model. $\otimes$ denotes the non-dominant user, and $\bullet$ denotes the dominant user.}%\vspace{-3mm}
        \label{Fig-geomodel}
\end{figure}

Based on Lemma 5.1 and the optimal solutions of Subproblems 5.1a and
5.1b, we are now able to obtain the following lemma, Lemma 5.2,
which will be used to prove Theorem 5.1. Proof of Lemma 5.2 requires
assuming the path loss model for the users' channel gains. That is,
the channel gain is given by $g=\alpha z^{-\gamma}$, where $\gamma =
2 $ is the path loss exponent, $z$ is the distance to either the AP
or the BS and $\alpha\ge 0$ is a constant factor. Consequently, the
results in Lemma 5.2 and Theorem 5.1 rely on the path loss model, a
geometry for the users, BS and AP and a probability distribution of
the users over the specified geometry. For the convenience of
exposition, we consider a 1 by 1 square area with the BS at
coordinate $(0,0)$ and the WiFi AP at $(1,1)$. We will assume that
the users are uniformly distributed. For simplicity, we give the
proof of Lemma 5.2 based on the geometry specified in
Fig.~\ref{Fig-geomodel}, but it is worth pointing out that the proof
extends to more general geometries with minor modifications.

% As noted in the Theorem, we use  Let $z^*$ be the solution to the equation $c z^{-\gamma} = \frac{\left(e-1\right)\sigma_A^2}{P}$. Let $z'\ge 0$ be the maximum value such that $\frac{2c}{(1-z')^2+1} \ge \frac{c}{(\sqrt{2}-z')^2}$. That is, $z' = 0.67$. Let $C$, as specified in Fig.~\ref{Fig-geomodel}, be $C = \min\{z^*, 0.67\}$. Before we prove Theorem 5.1, we first prove the following lemma.

\underline{\textbf{Lemma 5.2}}: Let $z^*$ be the solution to the
equation $\alpha z^{-\gamma} =
\frac{\left(e-1\right)\sigma_A^2}{P}$. Let $D$, as specified in
Fig.~\ref{Fig-geomodel}, be $D = \min\{z^*, 0.67\}$ (the constant,
0.67, is derived in the proof of Lemma 5.2). Suppose there is at
least one user in the quarter circle with radius $D$ and centered at
the AP, as shown in Fig. \ref{Fig-geomodel}, then at the optimal
solution to Problem 5.1, at most one user connects to the AP and all
the other users connect to the BS.

\emph{Proof:} We first consider subproblem 5.1b. Since there is at least one user in the stated quarter circle, the user with the strongest channel gain to the AP has a channel gain of at least $\frac{\left(e-1\right)\sigma_A^2}{P}$. From Lemma 4.1 part (ii), at the optimal solution to subproblem 5.1b, only the user with the strongest channel gain transmits. We denote the transmitting user as user $k^*$, and refer to the transmitting user as the \textit{dominant} user.

Next, returning back to Problem 5.1, let $\mathcal {S}^*$ be the set of users connected to the
WiFi AP under the optimal solution. Based on Lemma 5.1, all users in $\mathcal {S}^{*C}$ will connect to the BS. Let where $|\cdot|$ denote the
cardinality of a set. We now show that $|\mathcal {S}^*| \le 1$, where $|\cdot|$ by contradiction.

Suppose first that $|\mathcal {S}^*|=2$. We have two possible
cases:
\begin{itemize}
  \item Case 1: \emph{With a dominant user in $\mathcal {S}^*$}. That is, user $k^* \in \mathcal {S}^*$.
  In this case, if we assign the non-dominant user to the BS, the utility at the BS side $\lambda R_B^w(\boldsymbol{x})$ will
  increase. On the other hand, from Lemma 4.1 part (ii), the utility at the AP side is maximized when only user $k^*$ connects to the AP. Thus, by assigning the non-dominant user to the BS, we can also increase the utility at the AP side $(\lambda-\mu)R_B^o(\boldsymbol{y})$. Hence, the total utility of the operator increases if we assign only user $k^*$ to the AP, and the rest to the BS. This contradicts the assumption that $|\mathcal {S}^*| = 2$.
  \item Case 2: \emph{Dominant user not in $\mathcal {S}^*$}. That is, user $k^* \notin \mathcal {S}^*$. Denote the channel power gain of the channel between the dominant user and the BS
  as $h_{k^*,B}$, and the channel
  power gains of the channels between the two non-dominant users and the BS as
  $h_{m,B}$ and $h_{n,B}$, respectively. Now, consider the case where we switch the connections of $k^*$, and $m$ and $n$. In this case, the utility of the AP clearly increases by Lemma 4.1 part (ii), but the utility at the BS may not increase. However, it is straightforward to verify that the utility at the BS increases if the following condition holds.
\begin{align}\label{Eq-CHCON}
h_{m,B}+h_{n,B}\ge h_{k^*, B}.
\end{align}

 % \iffalse
%  Based on the relationship
%  between $h_{k^*, B}$ and $h_{m,B},~h_{n,B}$, it can be observed that
%  if we switch the two non-dominant users to the BS
%  and switch the dominant user to the WiFi AP, we have  four possible scenarios,
%  as listed in the following table.
%\begin{table}[!h]
%\tabcolsep 0pt \caption{Four Possible Scenarios} %\vspace*{-12pt}
%\begin{center}\label{Table-1}
%\begin{tabular}{|c|c|c|c|c|}
% \hline
% &~Conditions~& ~Utility at BS~ &~Utility at AP~ &~Total Utility~~\\
% \hline
% ~$1$~ & ~$h_{k^*, B}<h_{m,B},~h_{k^*, B}<h_{n,B}$~ &Increase &Increase&Increase\\
% \hline
%$2$ & ~$h_{k^*, B}<h_{m,B},~h_{k^*, B}>h_{n,B}$~ &Increase &Increase&Increase\\
% \hline
%$3$ & ~$h_{k^*, B}>h_{m,B},~h_{k^*, B}<h_{n,B}$~ &Increase &Increase&Increase\\
% \hline
%$4$ & ~$h_{k^*, B}>h_{m,B},~h_{k^*, B}>h_{n,B}$~ &Uncertain &Increase&Uncertain\\
% \hline
% \end{tabular}
% \end{center}
% \end{table}
%
%For the first three cases  the total utility of the cellular
%operator increases if we switch the connections of the non-dominant
%users with that of the dominant user. However, for the last case,
%though the utility at the AP side increases, it is not certain that
%if the utility at the BS side increases.
%Nonetheless, we observe
%that the utility at the BS side increases if the following
%condition holds:
%\begin{align}\label{Eq-CHCON}
%h_{m,B}+h_{n,B}\ge h_{k^*, B}.
%\end{align}
%\fi

Now, referring to Fig.~\ref{Fig-geomodel}, let the dominant user be a distance of $\sqrt{2} - d$ away from the BS, where $d \le D$ under the conditions stated in the Lemma. The two users in $\mathcal {S}^*$ have to be at least distance $d$ away from the AP, since their channel gains to the AP is weaker than the dominant user's. Considering the worst case scenario as given in Fig.~\ref{Fig-geomodel}, we have
\begin{align}
h_{m,B}+h_{n,B} &\ge \frac{2\alpha}{(1-d)^2+1} \nonumber\\
& \stackrel{a}{\ge} \frac{\alpha}{(\sqrt{2}-d)^2} \nonumber\\
& \ge h_{k^*, B},
\end{align}
where ``a'' follows from $d \le D \le 0.67$. Hence,
inequality~(\ref{Eq-CHCON}) holds under the conditions stated in
Lemma 5.2. Therefore, the total utility increases by switching the
two non-dominant users with the dominant user, which contradicts our
assumption that $|\mathcal {S}^*| = 2$.
\end{itemize}
Using the same arguments, we can show that any $|\mathcal {S}^*| >2$ results in a contradiction under the conditions stated in Lemma 5.2. Hence, $|\mathcal {S}^*| \le 1$, which completes the proof of Lemma 5.2. \hfill $\square$

We are now ready to prove Theorem 5.1.

\underline{\textbf{Theorem 5.1}}: When the number of users $N$ is
large, with high probability the optimal solution for Problem 5.1
under path loss model is: At most one user connects to the AP and
all the other users connect to the BS.

\emph{Proof:}
Since the users are uniformly distributed over the square of area one given in Fig.~\ref{Fig-geomodel}, the probability that there is at least one user in the quarter circle with radius $D$ and centered at the AP is $\mbox{Prob}(AP) = 1 - (1 - \pi D^2/4)^N$. Since $D > 0$, $\mbox{Prob}(AP) \to 1$ as $N \to \infty$. Hence, the condition in Lemma 5.2 holds with high probability, which implies that the assertion in Theorem 5.1 holds with high probability.\hfill $\square$

Based on the result given in Theorem 5.1, the optimal solution of
Problem 5.1 can be easily found by the following algorithm, which is
given in Table II.
\begin{table}[!h]
\tabcolsep 0pt \caption{}  \vspace*{-15pt}
\begin{center}
\def\temptablewidth{0.48\textwidth}
{\rule{\temptablewidth}{1pt}}
\begin{tabular*}{\temptablewidth}{@{\extracolsep{\fill}}l}
Proposed Data Offloading Scheme for Problem 5.1 \\ \hline 1.
\textbf{For} $k=1:N$;
\\~~~~~~\textbf{initialize} $\boldsymbol{x}=[1,1,\cdots,1]^T$; $\boldsymbol{y}=[0,0,\cdots,0]^T$;
\\~~~~~~\textbf{set} $\boldsymbol{x}(k)=0$; $\boldsymbol{y}(k)=1$;
\\~~~~~~\textbf{compute} $\boldsymbol{F}(k)=U(\boldsymbol{x},\boldsymbol{y})$;
\\~~~\textbf{end}
\\ 2. Find the optimal allocation and the maximum value of $\boldsymbol{F}$,
\\~~~$[F_{max}, \mbox{index}]=\max{\boldsymbol{F}}$;
\\ 3. Compare $F_{max}$ with the utility without offloading
$U(\boldsymbol{1}_N,\boldsymbol{0}_N)$.
\end{tabular*}
{\rule{\temptablewidth}{1pt}}
\end{center}
\end{table}
\vspace*{-8pt}

%
%\begin{align}
%\mathscr{P}=1-\left(1-{\rm
%Prob}\left\{d>\left(\frac{P}{(e-1)\sigma^2}\right)^\frac{1}{\gamma}\right\}\right)^N
%\end{align}
%
%

%
%Relaxation:
%\begin{align}
%\max_{\{x_i,y_i,\forall i\}} ~~&\lambda
%\ln\kern-0.5mm\left(\kern-0.5mm1\kern-0.5mm+\kern-0.5mm\sum_{i=1}^N
%\frac{g_{i,B}P_i}{\sigma_B^2}\kern-0.5mm\right)\kern-1mm+(\lambda-\kern-0.5mm\mu)\sum_{i=1}^N
%\kern-0.5mm\ln\kern-0.5mm\left(\kern-0.5mm1\kern-0.5mm+\kern-0.5mm
%\frac{g_{i,A}\tilde{P_i}}{\sum_{j=1,j\neq i}^N
%g_{j,A}\tilde{P_j}+\sigma_A^2}
%\kern-0.5mm\right),\\
%\mbox{s.t.}~~~~& 0\le P_i \le P, ~\forall i,\\
%& 0\le \tilde{P_i} \le P, ~\forall i,\\
%& 0\le P_i+\tilde{P_i} \le P, ~\forall i.
%\end{align}

%\begin{align}
%\max ~~&\lambda
%\ln\kern-0.5mm\left(\kern-0.5mm1\kern-0.5mm+\kern-0.5mm\sum_{i=1}^N
%\frac{(1-y_i)g_{i,B}P}{\sigma_B^2}\kern-0.5mm\right)\kern-1mm+(\lambda-\kern-0.5mm\mu)\sum_{i=1}^N
%\kern-0.5mm\ln\kern-0.5mm\left(\kern-0.5mm1\kern-0.5mm+\kern-0.5mm
%\frac{y_i g_{i,A}P}{\sum_{j=1,j\neq i}^N y_j g_{j,A}P+\sigma_A^2}
%\kern-0.5mm\right),\\
%\mbox{s.t.}~~~~& y_i \in \{0, 1\}, ~\forall i,
%\end{align}
%
%Relaxation
%\begin{align}
%\max_{P_i, ~\forall i} ~~&\lambda \ln\kern-0.5mm\left(\kern-0.5mm
%S\kern-0.5mm-\sum_{i=1}^N\frac{g_{i,B}P_i}{\sigma_B^2}\kern-0.5mm\right)\kern-1mm+(\lambda-\kern-0.5mm\mu)\sum_{i=1}^N
%\kern-0.5mm\ln\kern-0.5mm\left(\kern-0.5mm1\kern-0.5mm+\kern-0.5mm
%\frac{g_{i,A}P_i}{\sum_{j=1,j\neq i}^N g_{j,A}P_j+\sigma_A^2}
%\kern-0.5mm\right),\\
%\mbox{s.t.}~~~~& 0 \le P_i \le P, ~\forall i,
%\end{align}
%where $S=1+\sum_{i=1}^N \frac{g_{i,B}P}{\sigma_B^2}$.

\section{Related Scenarios} \label{Sec-extension}
\subsection{Benefit of SIC Decoders}
In this subsection, we investigate the role of SIC decoders in the
utility maximization of the cellular operator. We rigorously prove
that the SIC decoder is beneficial for the cellular operator in
terms of maximizing its utility.
%\begin{align}
%U^{ww}(\boldsymbol{x},\boldsymbol{y})&=U^{w}_B(\boldsymbol{x})+U^{w}_A(\boldsymbol{y})\\
%U^{oo}(\boldsymbol{x},\boldsymbol{y})&=U^{o}_B(\boldsymbol{x})+U^{o}_A(\boldsymbol{y})\\
%U^{wo}(\boldsymbol{x},\boldsymbol{y})&=U^{w}_B(\boldsymbol{x})+U^{o}_A(\boldsymbol{y})
%\end{align}

\underline{\textbf{Theorem 6.1}}: Let
$(\boldsymbol{x}^*,\boldsymbol{y}^*)$,
$(\hat{\boldsymbol{x}}^*,\hat{\boldsymbol{y}}^*)$,
$(\tilde{\boldsymbol{x}}^*,\tilde{\boldsymbol{y}}^*)$ be the optimal
solutions of Problem 3.1, 4.1, and 5.1, respectively. In general,
the following inequality always holds,
\begin{align}
U^{ww}(\boldsymbol{x}^*,\boldsymbol{y}^*)\ge
U^{wo}(\tilde{\boldsymbol{x}}^*,\tilde{\boldsymbol{y}}^*)\ge
U^{oo}(\hat{\boldsymbol{x}}^*,\hat{\boldsymbol{y}}^*).
\end{align}

\emph{Proof:} To prove Theorem 6.1, we first show that
$U^{ww}(\boldsymbol{x}^*,\boldsymbol{y}^*)\ge
U^{wo}(\tilde{\boldsymbol{x}}^*,\tilde{\boldsymbol{y}}^*)$. It can
be observed that $U^{ww}(\boldsymbol{x}^*,\boldsymbol{y}^*)\ge
U^{ww}(\tilde{\boldsymbol{x}}^*,\tilde{\boldsymbol{y}}^*)$. This is
due to the fact that
$(\tilde{\boldsymbol{x}}^*,\tilde{\boldsymbol{y}}^*)$ is a feasible
solution of Problem 3.1, while $(\boldsymbol{x}^*,\boldsymbol{y}^*)$
is the optimal solution of Problem 3.1. Thus, if we can show that
$U^{ww}(\tilde{\boldsymbol{x}}^*,\tilde{\boldsymbol{y}}^*)\ge
U^{wo}(\tilde{\boldsymbol{x}}^*,\tilde{\boldsymbol{y}}^*)$ holds,
$U^{ww}(\boldsymbol{x}^*,\boldsymbol{y}^*)\ge
U^{wo}(\tilde{\boldsymbol{x}}^*,\tilde{\boldsymbol{y}}^*)$ will
hold.  Since
$U^{ww}(\tilde{\boldsymbol{x}}^*,\tilde{\boldsymbol{y}}^*)=\lambda
R^{w}_B(\tilde{\boldsymbol{x}}^*)+(\lambda-\mu)R^{w}_A(\tilde{\boldsymbol{y}}^*)$
and
$U^{wo}(\tilde{\boldsymbol{x}}^*,\tilde{\boldsymbol{y}}^*)=\lambda
R^{w}_B(\tilde{\boldsymbol{x}}^*)+(\lambda-\mu)R^{o}_A(\tilde{\boldsymbol{y}}^*)$,
we only need to show that $R^{w}_A(\tilde{\boldsymbol{y}}^*) \ge
R^{o}_A(\tilde{\boldsymbol{y}}^*)$ always holds, which is presented
as below.

Assume that $K$ elements of $\tilde{\boldsymbol{y}}^*$ are equal to
$1$, where $K\in\{1,2,\cdots,N\}$. Then, it follows that
\begin{align}
R^{w}_A(\tilde{\boldsymbol{y}}^*)=&\ln\left(1+\sum_{i=1}^K\frac{g_{i,A}P}{\sigma_A^2}\right)\nonumber
\\=&\ln\left(\frac{\sigma_A^2+\sum_{i=1}^K g_{i,A}P}{\sigma_A^2}\right)\nonumber
\\=&\ln\left[\left(\frac{\sigma_A^2+\sum_{i=1}^K
g_{i,A}P}{\sigma_A^2+\sum_{i=2}^K
g_{i,A}P}\right)\left(\frac{\sigma_A^2+\sum_{i=2}^K
g_{i,A}P}{\sigma_A^2+\sum_{i=3}^K g_{i,A}P}\right)\right.\nonumber\\
&\left.\cdots\left(\frac{\sigma_A^2+\sum_{i=K}^K
g_{i,A}P}{\sigma_A^2}\right)\right]\nonumber
\\\stackrel{a}{=}&\sum_{j=1}^K
\ln\left(\frac{\sigma_A^2+\sum_{i=j}^K
g_{i,A}P}{\sigma_A^2+\sum_{i=j+1}^K g_{i,A}P}\right)\nonumber
\\=&\sum_{j=1}^K \ln\left(1+\frac{
g_{j,A}P}{\sigma_A^2+\sum_{i=j+1}^K g_{i,A}P}\right)\nonumber
\\\stackrel{b}{\ge}&\sum_{j=1}^K \ln\left(1+\frac{
g_{j,A}P}{\sigma_A^2+\sum_{i=1,i\neq j}^K
g_{i,A}P}\right)=R^{o}_A(\tilde{\boldsymbol{y}}^*),
\end{align}
where we introduce a dumb item $\sum_{i=K+1}^K g_{i,A}P=0$ in the
equality ``a'' for notational convenience. The inequality ``b''
follows from the fact that $\sum_{i=1,i\neq j}^K g_{i,A}P \ge
\sum_{i=j+1}^K g_{i,A}P, \forall j$.

Then, it is clear that $U^{ww}(\boldsymbol{x}^*,\boldsymbol{y}^*)\ge
U^{wo}(\tilde{\boldsymbol{x}}^*,\tilde{\boldsymbol{y}}^*)$ always
holds. Using the same approach, we can easily show that
$U^{wo}(\tilde{\boldsymbol{x}}^*,\tilde{\boldsymbol{y}}^*)\ge
U^{oo}(\hat{\boldsymbol{x}}^*,\hat{\boldsymbol{y}}^*)$ always holds.
Theorem 6.1 is thus proved. \hfill $\square$

From Theorem 6.1, it is observed that SIC decoder plays an important
role in the utility maximization of the cellular operator. It is
beneficial for the operator to equip the BS with SIC decoders in
terms of maximizing its utility.

%Relaxation:
%\begin{align}
%\max ~~&\lambda \sum_{i=1}^N
%\kern-0.5mm\ln\kern-0.5mm\left(\kern-0.5mm1\kern-0.5mm+\kern-0.5mm
%\frac{g_{i,B}P_i}{\sum_{j=1,j\neq i}^N g_{j,B}P_j+\sigma_B^2}
%\kern-0.5mm\right)\kern-1mm+(\lambda-\kern-0.5mm\mu)\sum_{i=1}^N
%\kern-0.5mm\ln\kern-0.5mm\left(\kern-0.5mm1\kern-0.5mm+\kern-0.5mm
%\frac{g_{i,A}\tilde{P_i}}{\sum_{j=1,j\neq i}^N
%g_{j,A}\tilde{P_j}+\sigma_A^2}
%\kern-0.5mm\right),\\
%\mbox{s.t.}~~~~& 0\le P_i \le P, ~\forall i,\\
%& 0\le \tilde{P_i} \le P, ~\forall i,\\
%& 0\le P_i+\tilde{P_i} \le P, ~\forall i.
%\end{align}

\subsection{Distributed Data Offloading}
In the previous sections, we have obtained the optimal data
offloading schemes for Problem 3.1, 4.1, and 5.1 when the number of
users is large. \textcolor[rgb]{0,0,0}{However, the proposed data
offloading schemes are centralized schemes, which needs the users to
send the user-to-AP and user-to-BS channel power gains to the BS,
and then the BS has to compute the optimal user association and
feedback the decisions to the users. For Problem 4.1 and 5.1, due to
the special structure of the problems, the proposed centralized
algorithms can find the optimal solution in polynomial time.
However, for Problem 3.1, due to the complexity of the problem, the
proposed algorithm puts a heavy computational burden on the BS.
Thus, to relieve the computational burden on BS and reduce the
overhead for CSI and decision transfer, in this section, we propose
a simple but highly efficient distributed data offloading scheme for
Problem 3.1, which is given in Table III.}

\begin{table}[!h]
\tabcolsep 0pt \caption{}  \vspace*{-15pt}
\begin{center}
\def\temptablewidth{0.48\textwidth}
{\rule{\temptablewidth}{1pt}}
\begin{tabular*}{\temptablewidth}{@{\extracolsep{\fill}}l}
Proposed Distributed Data Offloading Scheme for Problem 3.1 \\   \hline 1). The cellular operator broadcasts an offloading threshold $T$ to each user. \\
2). User $i$ computes its value of $\frac{
S_{i,B}}{S_{i,A}},~\forall i$. If $\frac{ S_{i,B}}{S_{i,A}}\ge T$,
it connects to the \\ BS; Otherwise, it connects to the WiFi AP.
\end{tabular*}
{\rule{\temptablewidth}{1pt}}
\end{center}
\end{table}

%\textcolor[rgb]{0,0,0}{It is observed from Table III that the BS
%does not have to collect the CSI from the users, and it only needs
%to broadcast a predetermined threshold $T$ to the users. Thus, the
%network overhead of the distributed algorithm is much lower than
%that of the centralized algorithm. Besides, it is also observed that
%each user only has to compute a ratio ($\frac{ S_{i,B}}{S_{i,A}}$
%for user $i$) to decide its association. While in the centralized
%algorithm, the BS has to solve an integer programming problem to
%decide the optimal association for each user. Thus, the computing
%complexity of the distributed algorithm is much lower than that of
%the centralized algorithm. However, it is worthy pointing out that
%the performance of the distributed algorithm is in general not as
%good as that of the centralized algorithm. The performance of the
%distributed algorithm greatly depends on the value of the threshold
%$T$.}

\textcolor[rgb]{0,0,0}{It is observed from Table III that the BS
does not have to collect the CSI from the users, and it only needs
to broadcast a predetermined threshold $T$ to the users. Thus, the
network overhead of the distributed algorithm is much lower than
that of the centralized algorithm. On the other hand, the
computational complexity of the distributed algorithm is much lower
than that of the centralized algorithm. For the centralized
algorithm, the BS has to solve a relaxed integer programming problem
to decide the optimal association for each user, whose worst-case
computational complexity is $O(N^3$) \cite{complexity}. While for
the distributed scheme, the computational complexity is $O(N)$,
since each user only has to compute a ratio ($\frac{
S_{i,B}}{S_{i,A}}$ for user $i$) to decide its association. However,
it is worth pointing out that the performance of the distributed
algorithm greatly depends on the value of the threshold $T$.}

In the following, we show that the distributed data offloading
scheme can achieve the same performance as the centralized one given
in Table I if the threshold $T$ is properly chosen.

\underline{\textbf{Theorem 6.2}}: There exists an optimal threshold
$T^*$, for any user $i$ other than the user with fractional
indicator function, the following equality holds.
\begin{align}
x_i^*=\left\{\begin{array}{cc}
               1, & \mbox{if}~\frac{
S_{i,B}}{S_{i,A}}>T^*, \\
               0, & \mbox{if}~\frac{
S_{i,B}}{S_{i,A}}<T^*,
             \end{array}
\right.
\end{align}
where
$T^*=\frac{(\lambda-\mu)}{\lambda}\frac{1\kern-0.5mm+\kern-0.5mm\sum_{i=1}^N
S_{i,B}x_i^*}{1\kern-0.5mm+\kern-0.5mm\sum_{i=1}^N
S_{i,A}(1-x_i^*)}$, and $x_i^*, \forall i$ is the optimal solution
of Problem 3.3.

\emph{Proof:}
This proof is based on the KKT conditions given out in Section
\ref{Sec-WBoth}. It is observed from \eqref{eq-KKT01} that if
$\frac{ S_{i,B}}{S_{i,A}}>T^*$, where
$T^*=\frac{(\lambda-\mu)}{\lambda}\frac{1\kern-0.5mm+\kern-0.5mm\sum_{i=1}^N
S_{i,B}x_i^*}{1\kern-0.5mm+\kern-0.5mm\sum_{i=1}^N
S_{i,A}(1-x_i^*)}$, it follows that $\alpha_i-\beta_i>0$. From
\eqref{eq-KKT02} and \eqref{eq-KKT03}, it is also observed that
$\alpha_i\neq 0$ and $\beta_i\neq 0$ can not hold simultaneously.
Since $\alpha_i$ and $\beta_i$ are nonnegative, thus if $\beta_i>0$,
$\alpha_i$ must be equal to zero. Consequently, we have
$\alpha_i-\beta_i<0$, which contradicts with the fact that
$\alpha_i-\beta_i>0$. Thus, it clear that $\alpha_i>0$ and
$\beta_i=0$. Then, from \eqref{eq-KKT02}, it follows that $x_i=1$.
Similarly, when $\frac{ S_{i,B}}{S_{i,A}}<T^*$, it can be shown that
$\alpha_i=0$ and $\beta_i>0$, which indicates that $x_i=0$.

Theorem 6.2 is thus proved. \hfill $\square$

\subsection{Fading Scenarios}
\label{Sec-fading scenario} \textcolor[rgb]{0,0,0}{In this paper, we
consider three cases: (1) With SIC decoders at both sides; (2)
Without SIC decoders at both sides; (3) With a SIC decoder at one
side. It is worth pointing out that we do not assume any specific
distribution of the channel power gains for Cases (1) and (2). Thus,
the results obtained for Cases (1) and (2) can be directly applied
to the block-fading scenario \cite{6}, where the channel remains
constant during each fading block but possibly changes from one
block to another. For the block-fading scenario, we can solve the
utility maximization problem for each fading block, and update the
user association scheme every fading block. This is due to the fact
that there are no coupling constraints between the fading blocks,
and thus maximizing the utility function for each fading block is
equivalent to maximizing the long-term utility function \cite{7},
i.e., $\mathbb{E}\left[U(\boldsymbol{x},\boldsymbol{y})\right]$,
where $U(\boldsymbol{x},\boldsymbol{y})$ is given by equation
\eqref{eq-utilityobj} and the expectation is taken over the
probability distribution of all the involved channel power gains.
Since we did not assume any specific distribution of the channel
power gains, the result holds for block-fading channels with any
fading distributions, such as Rayleigh fading, Rician fading,
Nakagami fading. However, for Case (3), we assumed the path loss
model when deriving the results. This is due to the following
reason. Using other fading channel models instead of the path loss
model makes the utility maximization problem for this case
mathematically intractable. Thus, the offloading scheme proposed for
this case may not be optimal if fading channel models are adopted.
However, according to the simulation results presented in Section
\ref{NumericalResults}, the offloading scheme proposed for this case
also works well when fading channel models are considered.}

\subsection{Downlink Scenarios} \textcolor[rgb]{0,0,0}{In this
paper, we focus on the uplink scenario. In this subsection, we show
how to extend the obtained results to the downlink scenario. For the
downlink scenario, the system model becomes broadcast channels. For
broadcast channels, there are usually two implementation ways:
\begin{itemize}
  \item Superposition coding with SIC. The transmitter encodes the
messages for all the receivers using superposition coding. Each
receiver decodes the received message using SIC. This case is
similar to the uplink scenario with SIC. If we assume both BS and AP
adopt this scheme, and both of them adopt equal power allocation,
the resultant utility maximization problem can be obtained by
letting $S_{i,B}=\frac{g_{i,B}P}{\sigma^2_{i,B}}$ and
$S_{i,A}=\frac{g_{i,A}P}{\sigma^2_{i,A}}$, $\forall i$ in Problem
3.1. Then, we can solve this problem using the same approach as
Problem 3.1 by introducing
$g_{i,B}^\prime\triangleq\frac{g_{i,B}}{\sigma^2_{i,B}}$ and
$g_{i,A}^\prime\triangleq\frac{g_{i,A}}{\sigma^2_{i,B}}$.
  \item Orthogonal schemes. If SIC decoders are not available at the
  receivers, for the broadcast channel, the transmitter will not
  encode the message for all users together. Instead, they will use
  orthogonal schemes, such as TDMA. For this case, the resultant
  user association is trivial, i.e., in each time slot, one user
  is selected to connect to the BS, and one user is selected to connect to the
  AP.
\end{itemize}}

\section{Numerical Results} \label{NumericalResults}
In this section, numerical results are provided to evaluate the
performance of the proposed data offloading schemes. %It is shown that the proposed
%data offloading schemes can achieve good system performance.

\subsection{Simulation Parameters} The simulation setup is as follows. We consider a 1 by 1
square area with the base station at coordinate $(0,0)$ and the WiFi
AP at $(1,1)$. The number of users is denoted by $N$, and these $N$
users are uniformly scattered in the square. For simplicity,  we
assume that the transmit power of each user is the same and given by
$1$. \textcolor[rgb]{0.00,0.00,0.00}{Unless otherwise stated}, the
path loss model is adopted to model the channel power gain. Let
$(\mbox{posx}_i, \mbox{posy}_i)$ denote the position of user $i$,
then the channel power gain between it and the BS can be modeled as
$g_{i,B}=(\sqrt{\mbox{posx}_i^2+ \mbox{posy}_i^2})^{-\gamma}$, where
$\gamma$ is the path loss coefficient. Similarly, the channel power
gain between user $i$ and the AP can be modeled as
$g_{i,A}=(\sqrt{(1-\mbox{posx}_i)^2+
(1-\mbox{posy}_i)^2})^{-\gamma}$. In this paper, we consider the
free space path loss model where $\gamma=2$. The power of the
additive Gaussian noises at the BS and the AP are set to $1$, i.e.,
$\sigma_B^2=1$ and $\sigma_A^2=1$. The revenue coefficient $\lambda$
of the BS is set to 1, and the cost coefficient $\mu$ is set to
$0.5$. Matlab is used for running all the simulations.

\subsection{With SIC Decoders at Both Sides}

\subsubsection{Performance of the Centralized Data Offloading Scheme}
\begin{figure}[t]
        \centering
        \includegraphics*[width=9cm]{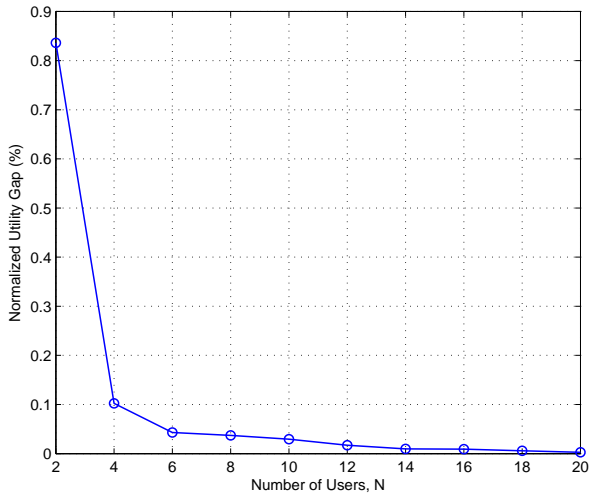}%\vspace{-3mm} %*[width=8cm]
        \caption{With SIC decoders at both sides: normalized utility gap vs. the number of users.}%\vspace{-3mm}
        \label{Fig-CentralizedData1}
\end{figure}
In Fig. \ref{Fig-CentralizedData1}, we investigate the gap between
the proposed centralized data offloading scheme given in Table I and
the optimal solution. The optimal solution is obtained by the
exhaustive search. For the purpose of illustration, the gap is
normalized by the utility of the optimal solution. The result
presented in Fig. \ref{Fig-CentralizedData1} is averaged over $1000$
channel realizations for each $N$. It is observed from Fig.
\ref{Fig-CentralizedData1} that the normalized utility gap decreases
with the increase of the number of users. When there are only two
users in the network, the normalized utility gap is as large as
$0.85\%$. When the number of users goes up to $16$, the normalized
utility gap is almost zero. This is in accordance with the results
presented in Theorem 3.2.

%In Fig. \ref{Fig-CentralizedData2}, we study the performance of the
%proposed centralized data offloading scheme. In this case, we also
%generate $100$ channel realizations for each $N$. We count the
%number of realizations, in which the normalized utility gap exceeds
%the given values. The given values actually indicate the accuracy of
%the proposed data offloading scheme. It is observed from Fig.
%\ref{Fig-CentralizedData2} that the number of realizations decreases
%with the increasing number of users for all the four given values,
%as expected. It is also observed from Fig.
%\ref{Fig-CentralizedData2} that when there are only two users in the
%network, almost $40$ realizations will result in the utility gap
%that is larger than $0.5\%$. However, when the number of users
%increases to $6$, the number of realizations that results in a
%utility gap larger than $0.5\%$ is zero. This indicates that when
%the system requirement is not too high, the proposed data offloading
%scheme will perform well even when the number of users is not large.
%It is also observed from Fig. \ref{Fig-CentralizedData2} that, even
%when the system requirement is very high, such as $0.05\%$, the
%proposed data offloading scheme can produce a satisfactory
%performance when there are more than $16$ users in the network.

\subsubsection{Performance of the Distributed Data Offloading Scheme}
\begin{figure}[t]
        \centering
        \includegraphics*[width=9cm]{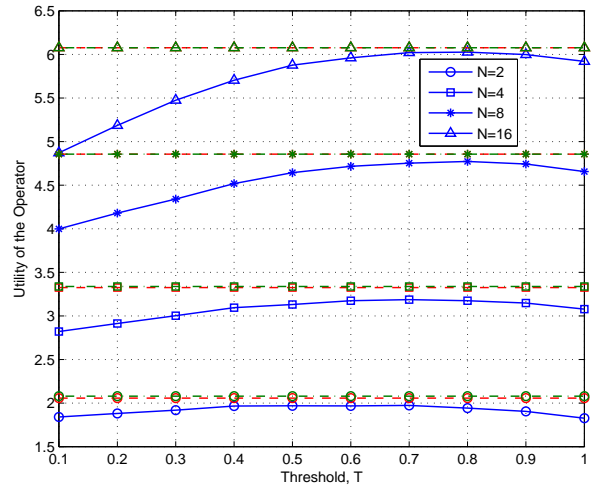}%\vspace{-3mm} %*[width=8cm]
        \caption{With SIC decoders at both sides: performance of the distributed data offloading scheme. }%\vspace{-3mm}
        \label{Fig-DistributedData}
\end{figure}
In Fig. \ref{Fig-DistributedData}, we investigate the system
performance of the proposed distributed data offloading scheme given
in Table III. The results presented in this figure is averaged over
$1000$ channel realizations. The red dashed lines represent the
values obtained by the centralized data offloading schemes. The
green dotted dashed lines represent the values obtained by the
exhaustive search. In this figure, we study how the value of the
threshold $T$ affects the performance of the proposed distributed
data offloading scheme.

It is observed from Fig. \ref{Fig-DistributedData} that the
centralized algorithm can achieve almost the same performance as the
exhaustive search, especially when $N$ is large. This is in
accordance with our theoretical results. It is observed that the
threshold $T$ plays a significant role in the distributed algorithm
when the number of users is large. It is observed that the utility
gap between the distributed algorithm and the exhaustive search is
as large as $1.2$ when $N=16$ if $T$ is not properly chosen.
However, when $N=2$, the largest utility gap is less than $0.2$. It
is also observed from Fig. \ref{Fig-DistributedData} that for each
$N$, there does exists an optimal $T$ which produces a utility which
is almost the same as the centralized data offloading scheme. This
is in accordance with the results presented in Theorem 6.2.

\subsection{Without SIC Decoders at Both Sides}
\begin{figure}[t]
        \centering
        \includegraphics*[width=9cm]{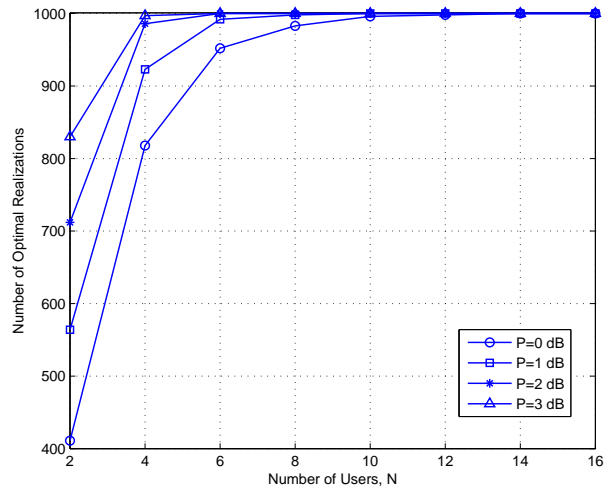}%\vspace{-3mm} %*[width=8cm]
        \caption{Without SIC decoder at both sides: performance of the data offloading scheme.}%\vspace{-3mm}
        \label{Fig-kxJ1Fig1}
\end{figure}
In Fig. \ref{Fig-kxJ1Fig1}, we investigate the performance of the
proposed data offloading scheme for the case that SIC decoders are
not available at both BS and the AP side. In Fig.
\ref{Fig-kxJ1Fig1}, we generate $1000$ channel realizations for each
$N$. We count the number of realizations, in which the proposed data
offloading scheme is optimal. First, it is observed that for all the
curves, the number of realizations that the proposed data offloading
scheme is optimal increases with the increasing number of users.
This is in accordance with our theoretical analysis given in Section
\ref{Sec-WOBoth}. Secondly, it is observed that the transmit power
of the users also plays an important role in the performance of the
proposed data offloading scheme. For the same number of users, when
the transmit power of the users is large, the number of realizations
that the proposed data offloading scheme is optimal is large. This
is due to the fact that when $P$ is large, the value of
$\frac{\left(e-1\right)\sigma_A^2}{P}$ is small, and thus the
probability that $g_{1,A}\ge \frac{\left(e-1\right)\sigma_A^2}{P}$
is large for the same number of users. Thirdly, it is observed that
when the number of users is larger than $10$, the proposed data
offloading scheme is always optimal for all the cases. This
indicates that the proposed data offloading scheme can achieve a
satisfactory performance even when the number of users is not very
large.

\subsection{With A SIC Decoder at One Side}
\begin{figure}[t]
        \centering
        \includegraphics*[width=9cm]{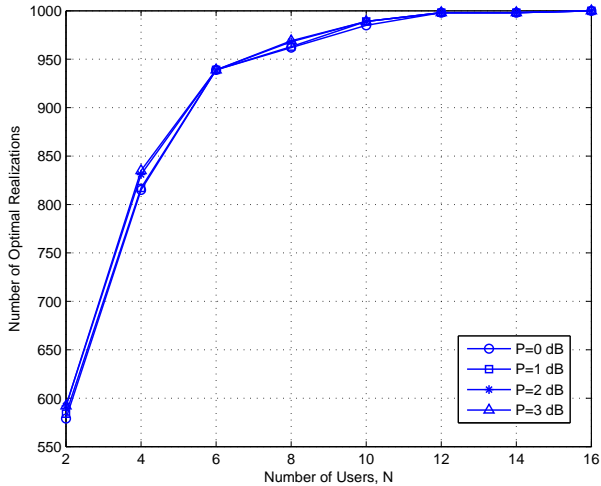}%\vspace{-3mm} %*[width=8cm]
        \caption{With SIC decoder at only BS: performance of the data offloading scheme.}%\vspace{-3mm}
        \label{Fig-kxJ1Fig2}
\end{figure}
In Fig. \ref{Fig-kxJ1Fig2}, we investigate the performance of the
proposed data offloading scheme for the case that a SIC decoder is
only available at the BS side. In Fig. \ref{Fig-kxJ1Fig2}, we
generate $1000$ channel realizations for each $N$. We count the
number of realizations, in which the proposed data offloading scheme
is optimal. It is observed that for all the curves, the number of
realizations that the proposed data offloading scheme is optimal
increases with the increasing number of users. This is in accordance
with our theoretical analysis in Section \ref{Sec-WOneSide}.
Secondly, it is observed that the transmit power of the users almost
does not affect the performance of the proposed data offloading
scheme. This is quite different from the results obtained in Fig.
\ref{Fig-kxJ1Fig1}. This is due to the fact that for this case, the
proposed data offloading scheme is optimal only when both
$g_{1,A}\ge \frac{\left(e-1\right)\sigma_A^2}{P}$ and $d<0.67$ are
satisfied simultaneously. For the case considered here, the
condition that $d<0.67$ always dominates. Since this condition is
irrelevant with the transmit power, the performance of the proposed
date offloading scheme is not affected by the transmit power of the
users. Finally, it is observed that when the number of users is
larger than $12$, the proposed data offloading scheme is always
optimal. This indicates that the proposed data offloading scheme can
achieve a good performance even when the number of users is not
large.

\begin{figure}[t]
        \centering
        \includegraphics*[width=9cm]{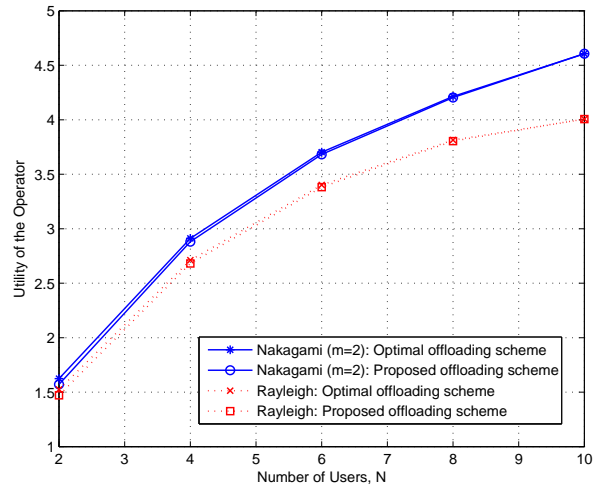}%\vspace{-3mm} %*[width=8cm]
        \caption{With SIC decoder at only BS: fading scenario.}%\vspace{-3mm}
        \label{Fig-case3fadingfig}
\end{figure}
\textcolor[rgb]{0,0,0}{In Fig. \ref{Fig-case3fadingfig}, we
investigate the performance of the proposed data offloading scheme
for different fading channel models. For the Rayleigh fading model,
the channel power gains are exponentially distributed
\cite{kangTWC}, and we assume that the mean of the channel power
gains is one.  For the Nakagami-m fading model, we consider the case
that $m = 2$, and we assume that the mean of the channel power gains
is one. The transmit power of each user is assumed to be the same
and equal to $1$. The results are averaged over 1000 channel
realizations. The optimal offloading schemes are obtained by
exhaustive search. It is observed from Fig. \ref{Fig-case3fadingfig}
that when the number of users is small, there is a small gap between
the proposed offloading scheme and the optimal offloading scheme.
However, when the number of users is larger than six, the proposed
offloading scheme can achieve same performance as the optimal
offloading scheme. This is due to the fact that when the number of
users is large, the condition given in \eqref{Eq-CHCON} holds with a
high probability, and thus the proposed offloading scheme is optimal
with a high probability. Overall, the proposed offloading scheme
works well under different fading channel models.}

%For the same number of
%users, when the transmit power of the users is large, the number of
%realizations that the proposed data offloading scheme is optimal is
%large. This is due to the fact that when $P$ is large, the value of
%$\frac{\left(e-1\right)\sigma_A^2}{P}$ is small, and thus the
%probability that $g_{1,A}\ge \frac{\left(e-1\right)\sigma_A^2}{P}$
%is large for the same number of users. Thirdly, it is observed that
%when the number of users is large than $10$, the proposed data
%offloading scheme is always optimal for all the cases. This
%indicates that the proposed data offloading scheme can achieve a
%satisfactory performance even when the number of users is not very
%large.

\subsection{Benefit of SIC Decoders}
\begin{figure}[t]
        \centering
        \includegraphics*[width=9cm]{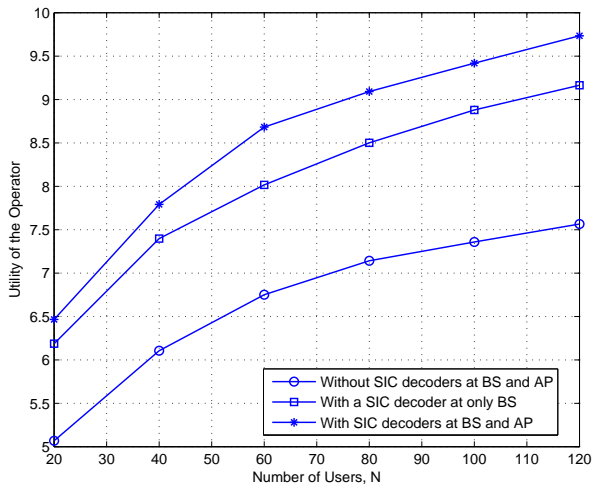}%\vspace{-3mm} %*[width=8cm]
        \caption{Benefit of the SIC decoders.}%\vspace{-3mm}
        \label{Fig-kxJ1Fig5}
\end{figure}
In Fig. \ref{Fig-kxJ1Fig5}, we compare the utility of the cellular
operator for the three cases studied in this paper. The utility
values for each case are obtained under their respective optimal
data offloading schemes.  The results presented in Fig.
\ref{Fig-kxJ1Fig5} are averaged over $1000$ channel realizations for
each $N$. It is observed that the utility increases with the
increasing of $N$ for all three cases. This is in accordance with
the theoretical results presented in previous sections. It is also
observed that $U^{ww}> U^{wo}> U^{oo}$ for the same $N$. This
indicates that SIC decoders have a significant effect on the utility
of the cellular operator. It is always beneficial for the operator
to equip the BS and/or AP with SIC decoders so as to maximize its
utility. This is in accordance with the results obtained in Theorem
6.1.

\section{Conclusions}\label{Sec-conclusion}
In this paper, we have investigated the mobile data offloading
problem through a third-party WiFi AP for a cellular mobile system.
From the cellular operator's perspective, we have formulated the
problem as a utility maximization problem. By considering whether
SIC decoders are available at the BS and/or the WiFi AP, different
cases are considered. When the SIC decoders are available at both
the BS and the WiFi AP, the utility maximization problem can be
solved by considering its relaxation problem. It is strictly proved
that the proposed data offloading scheme is near-optimal when the
number of users is large. We also propose a threshold-based
distributed data offloading scheme which can achieve the same
performance as the centralized data offloading scheme if the
threshold is properly chosen. When the SIC decoders are not
available at both the BS and the WiFi AP, we have rigorously proved
that the optimal solution is One-One-Association, i.e., one user
connects to the BS and the other user connects to the WiFi AP. When
the SIC decoder is only available at the BS, we have shown that
there is at most one user connecting to the WiFi AP, and all the
other users connect to the BS. We also have rigorously proved that
SIC decoders are beneficial for the cellular operator in terms of
maximizing its utility.

\section*{ACKNOWLEDGMENT}
\textcolor[rgb]{0.00,0.00,0.00}{We would like to express our sincere
thanks and appreciation to the associate editor and the anonymous
reviewers for their valuable comments and helpful suggestions. This
has resulted in a significantly improved manuscript.}
\end{document}